\documentclass[twocolumn,floatfix]{revtex4-2}

\usepackage[T1]{fontenc}
\usepackage{amsmath}
\usepackage{amssymb}
\usepackage{amsfonts,amsthm,bm}
\usepackage{xfrac}
\usepackage{nicematrix}
\usepackage[caption=false]{subfig}
\usepackage{mathrsfs}
\usepackage{tikz}
\usepackage{pgfplots}
\pgfplotsset{compat=1.18}
\usetikzlibrary{arrows.meta}
\usetikzlibrary{external}
\usepackage{standalone}
\usepackage[colorlinks=true]{hyperref}

\newcommand{\hamil}{\mathcal{H}}

\newcommand{\euler}[1]{\mathrm{e}^{#1}}

\renewcommand{\vec}{\mathbf}

\begin{document}

\begin{abstract}

Quantum geometry characterizes the variation of wavefunctions in momentum space through their overlaps and relative phases, providing a general framework for understanding many transport and optical properties. It is generally formulated in terms of interband matrix elements, which, entering the response functions, allow obtaining experimental access to the quantum geometric tensor. Recently, it has been emphasized that quantum geometry can also be interpreted in terms of quantum dipole fluctuations in the ground state driven by interband mixing. Here, we extend this picture to include contributions from many-body collective fluctuations, in which propagators and response vertices are dressed dynamically by the interaction with collective modes. Focusing on the Berry curvature, we show that contributions from collective fluctuations can be experimentally distinguished from bare band-geometric contributions, via specific antisymmetric channels in inelastic scattering spectra. We further identify the non-commutative properties of transverse quantum fluctuations as well as non-local-time interactions as the generators of this dynamical curvature in the susceptibility response. 
\end{abstract}

\title{Geometric curvature driven by many-body collective fluctuations}
\author{Alejandro S. Miñarro, Gervasi Herranz}
\affiliation{Institut de Ci\`encia de Materials de Barcelona (ICMAB-CSIC), Campus UAB, 08193 Bellaterra, Catalonia, Spain}

\maketitle

\section{INTRODUCTION}
\subsection{Quantum geometry and its interpretation in terms of quantum dipole fluctuations}

Geometry has long played a fundamental role in physics, from spacetime and gauge theories to condensed matter \cite{cheng2010quantum, liu2025quantum, yu2025quantum, gao2025quantum, torma2023essay}. In recent years, geometric concepts have experienced an important surge through the study of quantum geometry, as a unified framework describing a variety of remarkable phenomena in condensed matter systems \cite{jiang2025revealing, kim2025direct, mehraeen2025quantum, smith2022momentum}. For instance, broken inversion and/or time reversal symmetries can reveal Berry curvatures in nonequilibrium band populations, which generate nonlinear transport phenomena \cite{wang2024nonlinear, ortix2021nonlinear, kaplan2024unification, sala2025quantum}. On the other hand, quantum metric contributions are crucial to understand flatband superconductivity \cite{hase2025recent, classen2025high, huhtinen2022revisiting, torma2022superconductivity,adak2024tunable, chu2020superconductivity}, while the physics of fractional Chern insulators is driven by many-body electron interactions shaped by quantum geometry \cite{morales2024fractionalized, cao2025fractional, ju2024fractional, zhao2025exploring}. Recently, such concepts have been extended to optical phenomena, where nontrivial geometry may emerge in the space of optical transitions rather than Hilbert space \cite{ahn2022riemannian, nagaosa2024nonreciprocal, wu2025quantum, gao2019nonreciprocal, ahn2020low}. Thus, geometric curvature can arise near optical resonances, associated with higher-order responses or effective broken symmetries, giving rise to nonlinear optical and photovoltaic responses \cite{ahn2022riemannian, nagaosa2017concept, ma2023photocurrent, carmichael2025probing}. This has motivated the investigation of linear responses, where generalized sum rules and relations to the quantum metric have been established in longitudinal linear optical conductivity \cite{souza2000polarization,onishi2024fundamental, yu2025quantum, ghosh2024probing, wiedmann2025quantum}. 

\begin{figure}[t]
\centering
\includegraphics[width=\columnwidth]{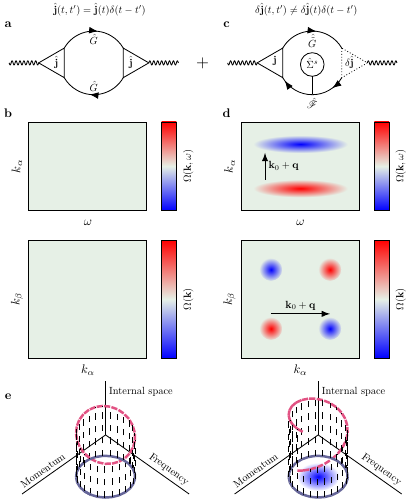}
\caption{\textbf{a} Susceptibility response in terms of a particle–hole bubble diagram with bare (instantaneous) current vertices $\hat{\vec{j}}_{\mathbf{k}}(t-t')= \hat{\vec{j}}_{\mathbf{k}}(t)\delta (t-t')$, with particle-hole lines corresponding to Green's functions $\hat{G}$ determined self-consistently. \textbf{b} In $\mathcal{P}$-$\mathcal{T}$ systems, the spectral and the geometrical Berry curvatures derived from this diagram vanish at all momenta and frequencies, $\Omega(\vec{k}, \omega)=0$, $\forall(\vec{k}, \omega)$ (section \ref{sec:linear_resp_A}). \textbf{c} Diagram corresponding to the first-order expansion of susceptibility. The coupling to fluctuations generates non-local-time vertex corrections $\hat{\vec{j}}_{\mathbf{k}}(t-t') \neq \hat{\vec{j}}_{\mathbf{k}}(t)\delta (t-t')$ and self-energy $\Sigma(\vec{k},\omega)$ with non-trivial momentum and frequency dependence, allowing nonzero spectral Berry curvature $\Omega(\vec{k}, \omega)\neq 0$. \textbf{d} The resulting spectral $\Omega(\mathbf{k},\omega)$ and geometric $\Omega({\vec{k}})$ curvatures exhibit localized hotspots in momentum-resolved maps (section \ref{sec:Berry_curv}). Inelastic probes with finite momentum transfer $\mathbf{q}$ can access transitions between regions of large spectral curvature (section \ref{sec:signatures_RIXS}). \textbf{e} Dynamical fluctuations induce an effective gauge, whereby closed loops in momentum-frequency space cause nontrivial complex rotations of the internal state. The emerging gauge is caused by the non-commutative properties of transverse fluctuation operators (section \ref{sec:linear_resp_B} and Figure \ref{fig:berryorigin}). Thus, both non-local-time interactions and non-commutative transverse fluctuations are key ingredients for the emergence of fluctuation-driven geometry.}
\label{fig:concept}
\end{figure}

It has recently been emphasized that this diversity of phenomena can be understood within the unifying framework of dipole fluctuations of the ground state emerging from interband mixing \cite{kohn1964theory,hetenyi2023fluctuations,onishi2024quantum, yu2025quantum,verma2025quantum, verma2026quantum}. Additionally, the quantum geometry encoded in the phase space of dipole fluctuations can be extracted from experiments exploiting response functions \cite{komissarov2024quantum,yu2025quantum, jiang2025revealing,gao2025quantum, verma2025quantum}. This provides a powerful route to analyze fundamental properties of quantum systems. For example, quantum information can be investigated by the analysis of the quantum metric, which is the symmetric real part of the quantum geometric tensor \cite{yu2022quantum, gao2025quantum, yu2025quantum}, which has recently been used to obtain information on spatial entanglement and charge fluctuations \cite{tam2024corner, wu2025corner} as well as the entanglement of orbital degrees of freedom \cite{ren2024witnessing}. However, these developments are generally formulated in terms of bare band-geometric descriptions \cite{verma2025quantum, verma2026quantum}. The incorporation of manybody effects, naturally extending the present framework, is therefore increasingly perceived as an important challenge \cite{yu2025quantum,jin2026experimental}. Thus, efforts have been aimed at revealing the quantum geometry of collective modes, e.g. magnon or spin excitations \cite{saji2025quantum, wu2026direct} or at relating manybody geometry to response and correlation functions \cite{guan2026exploring}. 

Here we address the following question: is there any specific response whose signal can selectively reveal genuine manybody effects in quantum geometry? To answer this question, we focus on the imaginary part of the quantum geometric tensor, the Berry curvature, which is more easily detected experimentally than quantum metric \cite{yu2025quantum, jiang2025revealing, kim2025direct}. The Berry curvature is related to a broad range of antisymmetric, Hall-like responses \cite{nagaosa2010anomalous, xiao2010berry}, thus providing a way for their experimental detection. To find a susceptibility channel that exclusively probes contributions from fluctuations, we impose specific symmetry conditions. In particular, we exploit the fact that, as long as we consider only band-geometric contributions, the antisymmetric off-diagonal susceptibility response should vanish identically in inversion ($\mathcal{P}$) and time-reversed ($\mathcal{T}$) symmetric systems, consistent with Onsager reciprocity relations \cite{fried2014relationship}. Crucially, despite these symmetry constraints, we show that fluctuations allow a finite localized spectral curvature $\Omega(\vec{k},\omega)$, provided that it vanishes when integrated throughout the Brillouin zone, $\Omega(\mathbf{k},\omega) = -\Omega(-\mathbf{k},\omega) \;\Rightarrow\; \int_{\mathrm{BZ}} \frac{d^d k}{(2\pi)^d}\,\Omega(\mathbf{k},\omega) = 0$. While optical spectroscopies are not suitable to experimentally detect these regions, here we show that inelastic spectroscopies, involving momentum transfer, can give access to this fluctuation-driven geometry.

\subsection{Geometric curvature driven by dynamical fluctuations}
We analyze how the interaction with a fluctuating field provides a distinct mechanism to generate spectral $\Omega(\mathbf{k},\omega)$ and geometric $\Omega({\mathbf{k}})$ curvatures in momentum-frequency space. Ignoring fluctuations, the effective momentum gauge structure in $\mathcal{P}-\mathcal{T}$ systems is trivial and the curvature vanishes identically in the response function. To fully isolate the geometric contribution from fluctuations, we first define a reference response by describing the linear susceptibility in terms of instantaneous current vertices $j_\vec{k}(t,t')\propto\delta(t-t')$ connected by two fermionic lines, corresponding to the bare (undressed) susceptibility bubble diagram shown in Figure \ref{fig:concept}a. Because of the constraints imposed by $\mathcal{P}-\mathcal{T}$ symmetries, both geometric and optical spectral curvatures vanish throughout the $(\vec{k}, \omega)$ space of the susceptibility response (except for singularities associated with isolated band degeneracies where it is ill-defined, such as in Dirac nodes at specific momenta, see Figure \ref{fig:concept}b).

A different situation, discussed more in depth in section \ref{sec:linear_resp_B}, arises as soon as coupling to a fluctuating field is considered, leading to a dressing of propagators and vertices and a distinctive dynamical curvature. In this situation, the interaction with collective fluctuations can generate a self-energy $\Sigma(\vec{k}, \omega)$ with nontrivial momentum and frequency structure (Figure \ref{fig:concept}c), allowing localized regions of finite curvature in the linear response function (Figure \ref{fig:concept}d). Crucially, the exchange mediated by bosonic propagators provides momentum-dependent vertex corrections $\delta j_\vec{k}(t,t')\not\propto\delta(t-t')$ that introduce time-nonlocality and memory effects in the effective interactions, driving nontrivial geometry in $(\vec{k}, \omega)$ space. This dynamically generated curvature requires all of the following ingredients: (i) a nontrivial momentum-frequency structure that gives rise to non-commutative parallel transport in $(\vec{k}, \omega)$ space due to the emergence of a fluctuation-driven gauge (\ref{fig:concept}e); (ii) a multiband structure that can host nontrivial internal phases; and (iii) a mechanism that generates such phases. Spin-orbit coupling is the most natural mechanism in $\mathcal{P}-\mathcal{T}$ symmetric systems, since it can create internal complex phases without breaking time-reversal symmetry.

\begin{figure}[tbh]
    \centering
    \includegraphics[width=0.8\columnwidth]{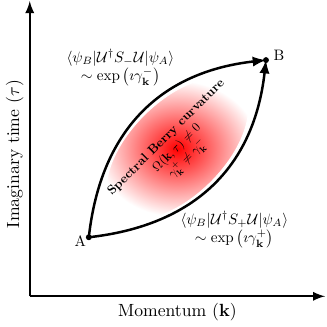}
    \caption{
In addition to time-non-local interactions (Figure \ref{fig:concept}), the origin of the geometric curvature generated by coupling to fluctuations is also related to the non-commutative properties of transverse quantum fluctuation operators and their propagation. We illustrate such effects in the plane of momentum and imaginary time, where spin ladder operators $S_+$, $S_-$ are propagated between two endpoints ($\vec{k}_1, \tau_1$), ($\vec{k}_2, \tau_2$), where $\mathcal{U}(\tau_2- \tau_1)$ is the propagator in momentum-(imaginary)time space. Due to the non-commutative characters of the operators, a geometric phase is accumulated between the two paths, which, together with nonlocal time interactions, is at the origin of the observed dynamical Berry curvature.
}
    \label{fig:berryorigin}
\end{figure}

The emergence of this effective gauge structure can be understood intuitively in terms of the non-commutative properties of the operators governing the propagation of quantum fluctuations. In Figure \ref{fig:berryorigin} we illustrate this point by considering ladder operators for transverse spin fluctuations $S_+, S_-$ applied in the space of momenta and imaginary times. In this context, the curvature arises from differences in the geometric phases accumulated when the fluctuation operators are propagated between two endpoints. Along each path in the $(\vec{k},\tau)$ manifold, the transverse quantum operators $S_+$ versus $S_-$ cannot diagonalize on the same spin-orbital basis, resulting in a nontrivial phase mismatch that can be interpreted as the flux of an enclosed spectral curvature. Thus, the geometric structure does not arise from propagation alone, but from the interplay between time-nonlocal propagation and the non-commutative character of the transverse fluctuation operators. Interestingly, this observation is consistent with the interpretation of anomalous Hall velocities and Berry curvature in terms of non-commuting effective position operators \cite{ishizuka2017noncommutative, xu2025quantum, gao2025quantum}. The role of time-nonlocality and non-commutativity in the emergence of dynamical curvature is developed in section \ref{sec:linear_resp_B}.

In Section \ref{sec:linear_resp_A}, we start by discussing the linear response of $\mathcal{P}-\mathcal{T}$ symmetric systems based on $t_{2g}$ states. The latter are relevant to describe the properties of a large family of heavy transition metal compounds that display strongly correlated states and large strong-spin-orbit coupling, which favors the emergence of unconventional phases \cite{takayama2021spin, khomskii2020orbital, brzezicki2020spin, matsuda2025kitaev, picozzi2024spin, natori2026ising}. Due to the imposed space inversion and time reversal symmetries, the antisymmetric transverse response is forced to vanish. In Section \ref{sec:linear_resp_B}, focusing on spin fluctuations, we show that an antisymmetric transverse component emerges in the response function due to the retardation effects and momentum dependence arising from coupling to collective modes. In Section \ref{sec:Berry_curv}, we interpret this antisymmetric response in terms of a geometric spectral curvature driven dynamically by fluctuations. Finally, in Section \ref{sec:signatures_RIXS}, we predict how resonance inelastic x-ray spectroscopy enables selective experimental access to this dynamical curvature, excluding bare band-geometric contributions. 

\section{LINEAR RESPONSE IN PARITY- AND TIME-REVERSAL-SYMMETRIC SYSTEMS} 
\subsection{Response from time-local vertices} \label{sec:linear_resp_A}

We first discuss the linear response in the first order of the diagrammatic expansion. This term, shown in Figure \ref{fig:concept}a, corresponds to the particle-hole bubble calculated from the paramagnetic current-current correlator in imaginary time $\chi_{ij}(\tau) = \left\langle \hat{j}_i(\tau)\hat{j}_j(0) \right\rangle$, where the current operator $\hat{j}_i = \sum_\vec{k} \hat{\vec{c}}_\vec{k}^\dagger \hat{j}^i_\vec{k} \hat{\vec{c}}_\vec{k}$ is built from bare (band-structure) time-local current vertices $\hat{j}^i_\mathbf{k}=e\partial_{k_i}\hat{H}(\mathbf{k})$. Note that, since we are interested in the finite-frequency response, we do not include here the diamagnetic contribution \cite{coleman2015introduction}, see also Appendix \ref{append_densitycurrent}. Therefore, the optical susceptibility is evaluated as

\begin{equation} \label{eq:bare}
\chi_{ij}(\vec{k},\tau)
=
-\sum_{\mathbf k}
\operatorname{Tr}\!\left[
\hat j^i_{{\vec{k}}}
\hat{G}_{\mathbf k}(\tau)\,
\hat j^j_{{\vec{k}}} \hat{G}_{\mathbf k}(-\tau)
\right]
\end{equation}
where the Green's functions $\hat{G}_{\mathbf k}(\tau)$ are determined self-consistently. We focus our study on strongly spin-orbit-coupled systems, typical of $4d-5d$ transition metal compounds, where the low-energy physics is described by $t_{2g}$ states \cite{witczak2014arcmp, khomskii2020orbital, takayama2021jpsj, pourovskii2025hidden}. For that, we use a Mastubara formalism, including intersite hopping, spin-orbit coupling, and Jahn-Teller interactions (Appendix \ref{append_latticemodel}). We take electron correlations at the second-order Born approximation, thus capturing dynamical effects in the weak-to-intermediate correlation regime. We then obtain the imaginary part of the retarded susceptibility $\Im\chi^{R}_{ij}(\mathbf{k},\omega)$ from the  spectral (Lehmann) representation $\chi_{ij}(\mathbf{k},\tau)=-\int_{-\infty}^{\infty} d\omega\,K(\tau,\omega)\,\frac{1}{\pi}\Im\chi^{R}_{ij}(\mathbf{k},\omega)$, where $K(\tau,\omega)=\frac{e^{-\omega\tau}}{1-e^{-\beta\omega}}$ is the analytic continuation kernel \cite{chikano2018irbasis, shinaoka2021efficient, wallerberger2023sparse}. Using the relation $\Im\chi^{R}_{ij}(\mathbf{k},\omega)=\omega\,\Re\sigma_{ij}(\mathbf{k},\omega)$, we determine the real part of the longitudinal conductivity \( \Re[\sigma_{xx}(\omega)] \).

\begin{figure*}[t]     
    \centering 
    \includegraphics[width=0.9\textwidth]{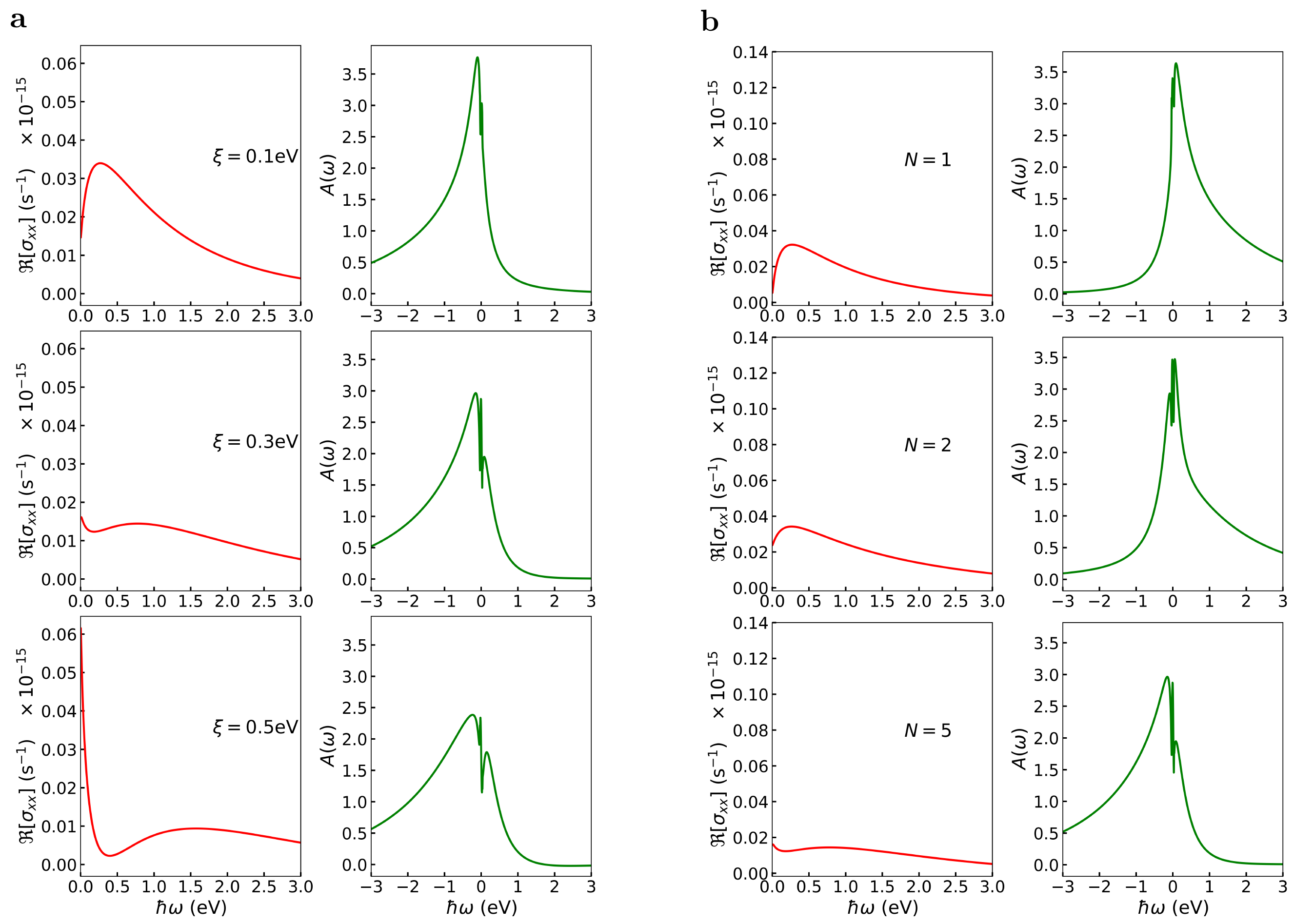}
    \caption{
(a) Real part of the optical conductivity \(\mathfrak{R}\left[\sigma_{xx}(\omega)\right]\) (left column) and the spectral function $\mathcal{A}(\omega) = -\frac{1}{\pi}\operatorname{Im} G^{R}(\omega)$ (right column) at orbital filling \(N=5\) of the $t_{2g}$ manifold, hopping \(t=0.2\,\text{eV}\), and increasing spin–orbit coupling (SOC, \(\xi\)). The conductivity is shown in Gaussian units s\(^{-1}\). For increasing SOC, the overall conductivity decreases, consistent with SOC-assisted reduction of kinetic energy, while an incoherent Hubbard band develops at higher energies. The emergence of a quasi-particle peak at low energies indicates weak-to-intermediate correlations, where dynamical effects are captured in the second-order Born approximation in our model. (b) The optical conductivity \(\mathfrak{R}\left[\sigma_{xx}(\omega)\right]\) and spectral function $\mathcal{A}(\omega)$ are plotted for orbital occupancies $N = 1,2$ and $5$, computed at fixed spin-orbit coupling $\xi = 0.3$ eV. The shape of $\mathcal{A}(\omega)$ reflects the expected dependence on the orbital filling of the $t_{2g}$ manifold. At low occupation ($N = 1$), electron removal (energies below the Fermi level $E_F$) is weak due to single occupancy of the lower-energy $J = 3/2$ quartet, while electron addition (energies above $E_F$) is large because of the available unoccupied states in both the quartet and higher-energy $J = 1/2$ doublet. In contrast, the situation is reversed at $N = 5$, where there stronger weight is expected at energies below $E_F$. The spectral function $\mathcal{A}(\omega)$ at $N = 2$ reflects a more balanced intermediate state.
}
\label{fig:longsigma}
\end{figure*}

 Figure~\ref{fig:longsigma}a shows \( \Re[\sigma_{xx}(\omega)] \) for orbital filling \( N = 5 \) and fixed hopping amplitude \( t = 0.2\,\text{eV} \). At low frequencies $(\omega \rightarrow 0)$ we observe a finite \( \Re[\sigma_{xx}(\omega)] \) spectral weight, which at larger spin-orbit coupling evolves into a quasi-particle peak, signaling a contribution from coherent transport. At the same time, as the spin-orbit coupling increases, some spectral weight is transferred to higher energies due to incoherent transport, signaling the emergence of Hubbard sidebands, whereas the overall conductivity decreases. Both observations indicate that spin-orbit coupling promotes the reduction of coherent transport, while enhancing correlations~\cite{georges1996rmp,laad2024aqt, kim2008novel, jackeli2009prl, kim2009phase, liu2011electronic, kim2012magnetic}, although the finite spectral weight at low $\omega$ suggests a moderate correlation regime, consistent with our perturbative approach to correlations. On the other hand, Figure~\ref{fig:longsigma}b shows \( \Re[\sigma_{xx}(\omega)] \) computed at fixed spin-orbit coupling $\xi = 0.3$ eV for different orbital occupancies. The data show a systematic redistribution of the spectral weight in the longitudinal response, being larger at low occupancy ($N = 1,2$), and significantly reduced at $N = 5$, reflecting the expected reduced phase-space availability for induced optical transitions in the latter.

On the other hand, the spectral functions derived from the retarded Green's function $\mathcal{A}(\omega) = -\frac{1}{\pi}\operatorname{Im} G^{R}(\omega)$ reflect the behavior expected for the different orbital occupations (Figure~\ref{fig:longsigma}). At low orbital filling ($N = 1$), the spectral weight below the Fermi level $E_F$ ($\omega < 0$) is significantly smaller than above $E_F$ ($\omega > 0$). This is in agreement with the single occupation of the lower-energy quartet $J = 3/2$ of the $t_{2g}$ manifold, which allows larger availability for transitions involving electron addition ($\omega > 0$) than removal ($\omega < 0$). The situation is reversed for $N = 5$, where the $J = 3/2$ quartet is fully occupied, while the $J = 1/2$ doublet has single occupancy. In this case, electron removal ($\omega < 0$) has larger spectral weight than electron addition ($\omega > 0$).

We now discuss the antisymmetric component of the susceptibility $\chi^{\mathrm{as}}_{ij}(\vec{k},\omega)$. We first note that an explicit analysis shows that in $\mathcal{P}-\mathcal{T}$ symmetric systems $\chi^{\mathrm{as}}_{ij}(\vec{k},\omega)$ vanishes identically for all frequencies and momenta, which implies the collapse of the optical spectral curvature throughout the phase space, i.e., $\Omega^{\alpha\beta}(\mathbf{k},\omega)
\propto \operatorname{Im}\,\chi^{\mathrm{as}}_{\alpha\beta}(\mathbf{k},\omega)=0$, $\ \forall(\vec{k},\omega)$. This observation can be understood from the fact that the Berry curvature satisfies \(\Omega^{\alpha\beta}(\mathbf{k},\omega)=-\Omega^{\alpha\beta}(-\mathbf{k},\omega)\) and \(\Omega^{\alpha\beta}(\mathbf{k},\omega)=\Omega^{\alpha\beta}(-\mathbf{k},\omega)\) due to the simultaneous presence of $\mathcal{P}$ and $\mathcal{T}$ symmetries. Therefore, in fully symmetric systems, the optical curvature vanishes throughout the Brillouin zone (except for isolated band degeneracies at specific momenta where it is ill defined, such as Dirac nodes), as long as only instantaneous (local-time) vertices are considered.

\begin{figure*}[t]
    \centering
    \includegraphics[width=0.9\textwidth]{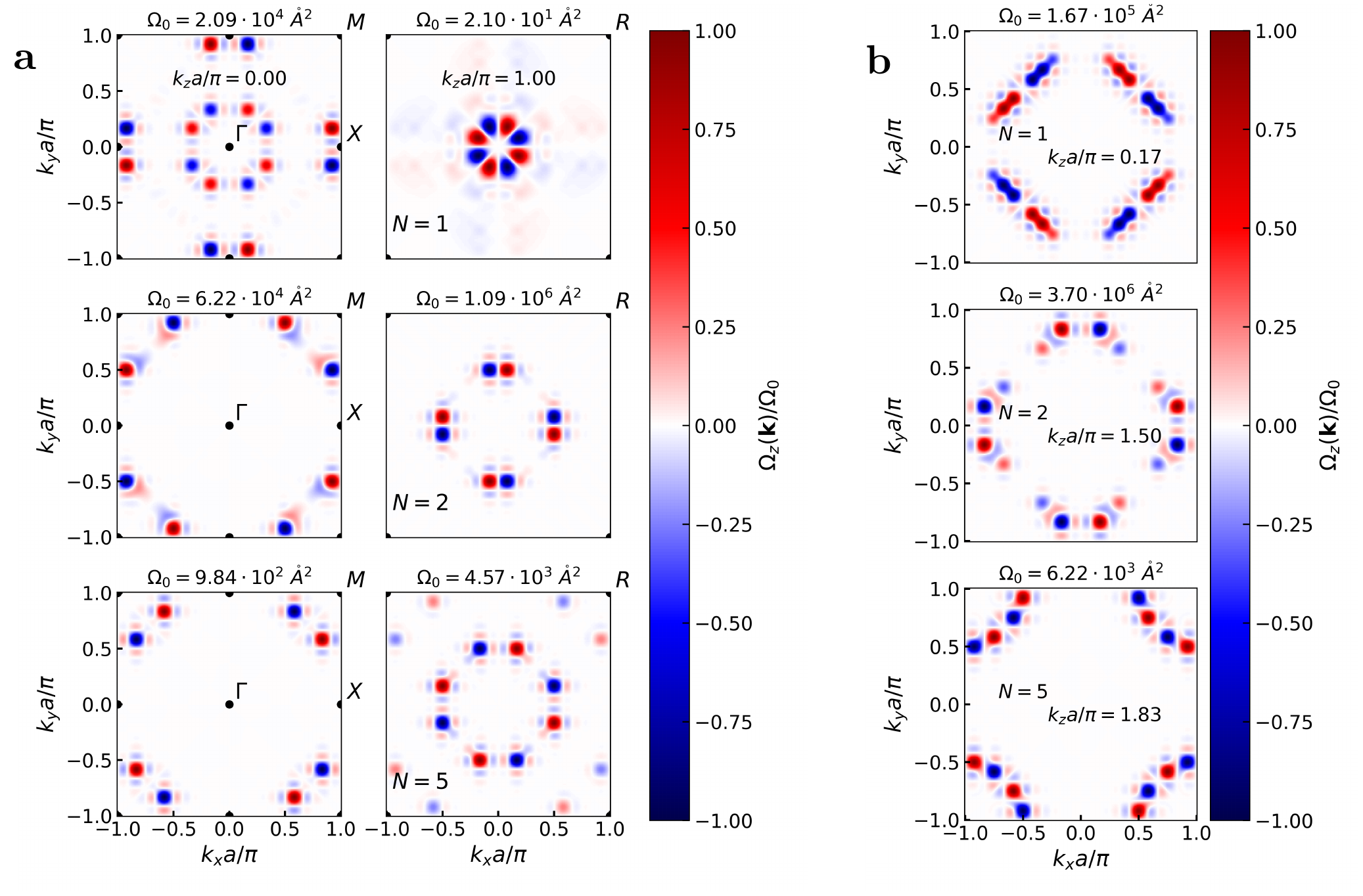}
    \caption{
(a) The geometric Berry curvature, given by $\Omega_z(\vec{k})=\int_{\omega\geq 0} d\omega \ \Omega_z(\vec{k},\omega)$ (Equation \ref{eq:dressedberry}), is mapped in the $k_x- k_y$ plane of the Brillouin zone, with $k_z=0$ (left column) and $k_z=\frac{\pi}{a}$ planes (right column). In this calculation, we assume anisotropic intersite hopping amplitudes $t_z \neq t_x=t_y$. The curvature is localized in specific regions with positive and negative values which, integrated over momenta, cancel. The maps are obtained at different orbital occupations of the $t_{2g}$ manifold $N=1,2,5$. The positions of high-symmetry points $\Gamma$, $X$, $M$, and $R$ are also indicated. (b) The curvature is mapped in the $k_x- k_y$ plane at values of $k_z$ where the its value is maximum. Data are normalized to the maximum $\Omega_0$ value in each panel.
}
    \label{fig:berrymaps}
\end{figure*}

\subsection{Dynamical curvature from time-nonlocal vertices and non-commutative fluctuation operators} \label{sec:linear_resp_B}
We now show that a finite spectral curvature emerges as soon as retarded (time-nonlocal) contributions and non-commutativity of quantum transverse fluctuations are taken into account. An intuition into these mechanisms is gained from the formulation of Hall-like responses in the Ishikawa-Mastuyama model for dc transport \cite{ishikawa1987microscopic,  zheng2019interaction, markov2021local}, in which the zero-frequency Hall conductivity is given by $\sigma_{xy} \propto \int d\omega\sum_{\mathbf{k}}\operatorname{Tr}\!\left[\hat{G}_{\mathbf{k}}\partial_{\omega} \hat{G}_{\mathbf{k}}^{-1}\, \hat{G}_{\mathbf{k}}\partial_{k_x} \hat{G}_{\mathbf{k}}^{-1}\, \hat{G}_{\mathbf{k}}\partial_{k_y} \hat{G}_{\mathbf{k}}^{-1}\right]$, which requires Green's functions with momentum and frequency dispersions.
This is reminiscent of the definition of the quantum geometric tensor $\hat{Q}_{\mu\nu} = \hat{P}\, \partial_{\mu} \hat{P}\, \partial_{\nu} \hat{P}$, which accounts for variations of quantum states along $\hat{\mu}$, $\hat{\nu}$ in momentum space projected by $\hat{P}$ onto a given subspace of Hilbert space \cite{torma2022superconductivity, yu2025quantum, liu2025quantum}. Similarly, coupling to dynamical fluctuations introduces non-commutativity and momentum and frequency dispersion via nonlocal time responses that allow frequency-dependent vertex corrections and finite curvature in some regions of the $(\vec{k},\omega)$ space.

To illustrate these points, we consider the first-order expansion of the bubble diagram that couples a fermionic line to a dynamic spin fluctuation (Figure \ref{fig:concept}c). Choosing orbital (instead of spin fluctuations) also gives a finite spectral curvature, though the details of the resultant geometric structure may vary. We then consider the following expression 
\begin{equation*}
\begin{split}
    \hat{\psi}^\alpha_\vec{k}(\tau) &= \psi_0 \mathrm{tr}\left[ \hat{\sigma}^\alpha \hat{G}_\vec{k}(\tau) \hat{\sigma}^\alpha \hat{G}_\vec{k}(-\tau) \right] \hat{\sigma}^\alpha  \\ & = \psi^\alpha_\vec{k}(\tau) \hat{\sigma}^\alpha
    \end{split}
\end{equation*}
where $\hat{\sigma}^\alpha$ represent spin Pauli matrices and $\psi_0$ is the coupling factor. As a result, a dynamical self-energy arises through the coupling of the dressed vertex $\hat{\psi}^\alpha_\vec{k}(\tau)$ to the fermionic propagator
\begin{equation*} \label{eq:sigmadynamic}
\begin{split}
       \hat{\Sigma}^s_\vec{k}(\tau) &= \dfrac{1}{\mathcal{N}} \sum_{\vec{q},\alpha} \hat{G}_{\vec{k}-\vec{q}}(\tau) \hat{\psi}^\alpha_\vec{q}(\tau).
\end{split}
\end{equation*}

To preserve the Ward-Takahashi identity associated with charge conservation \cite{mahan2000many, coleman2015introduction}, this self-energy contributes to a dressed current vertex
\begin{equation*}
\begin{split}
   \delta\hat{\vec{j}}_\vec{k}(\tau) &= -\dfrac{e}{\hbar} \boldsymbol{\nabla}_\vec{k} \hat{\Sigma}^s_\vec{k}(\tau).
\end{split}
\end{equation*}

\begin{figure*}[t]
    \centering
    \includegraphics[width=0.7\textwidth]{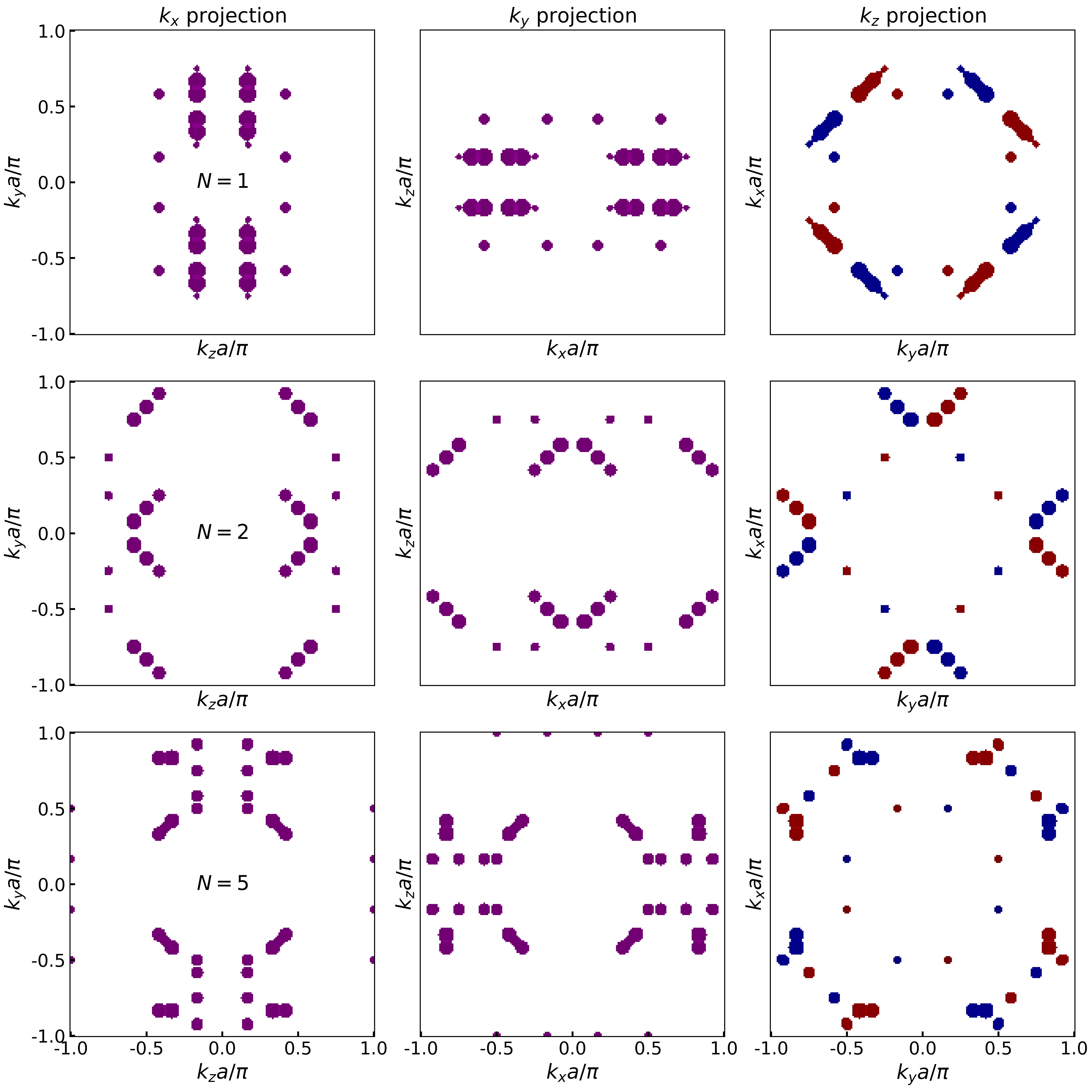}
    \caption{
The geometric Berry curvature $\Omega_z(\vec{k})= \int_{\omega\geq 0} d\omega \ \Omega_z(\vec{k},\omega)$ (assuming anisotropic intersite hopping amplitudes $t_z \neq t_x=t_y$) is mapped by projecting its values along different orientations, perpendicular to different planes: $k_y-k_z$ (left column), $k_x-k_z$ (middle) and $k_x-k_y$ (right). These projections show that the regions with finite curvature are localized in momentum space along all orientations. Purple color in the regions of finite curvature corresponds to overlapping regions with positive (red) and negative (blue) curvatures. The maps are obtained for different orbital occupations of the $t_{2g}$ manifold, $N=1$ (top row), $N=2$ (middle) and $N=5$ (bottom).     
}
    \label{fig:berryprojections}
\end{figure*}

In first loop approximation, this vertex correction can be expressed as
\begin{align*}
\delta\hat{\vec{j}}_{\vec{k}}(\tau)
&= \dfrac{j_0}{\beta\mathcal{N}}
\sum_{\vec{q},\alpha,\alpha',n}
\euler{-\imath\omega_n\tau}
\upsilon^\alpha \sin(k_\alpha a) \\
&\quad\times
\hat{G}_{\vec{k}-\vec{q}}(\imath\omega_n)
\hat{G}_{\vec{k}-\vec{q}}(\imath\omega_n)
\hat{\psi}^{\alpha'}_{\vec{q}}(\tau)\,
\vec{e}_\alpha
\end{align*}
where $\beta$ is the inverse temperature, $\vec{q}$ the momentum transfer, $\mathcal{N}$ the number of lattice sites, $\upsilon^{\alpha}=\delta t^\alpha/t$ (defined in Eq(\ref{app:eq:deltat})) the band velocity, $\imath\omega_n$ are Matsubara frequencies, and $j_0 = 2tea/\hbar$, where $t$ is the hopping amplitude, $e$ the electron charge, $a$ the lattice parameter, $\hbar$ the Planck constant, and $\vec{e}_\alpha$ is a unit vector in space. The antisymmetric off-diagonal optical susceptibility is then described by (see Appendix \ref{append_nonlocaltimecorrels})
\begin{equation*}
\begin{aligned}
\chi^{\mathrm{as}}_{\alpha\beta}(\tau)
&=
\sum_{\vec{k}}
\chi^{\mathrm{as}}_{\alpha\beta}(\vec{k},\tau),
\\[4pt]
\chi^{\mathrm{as}}_{\alpha\beta}(\vec{k},\tau)
&\equiv
-\mathrm{tr}\!\left\{
\left[
\hat{\Gamma}^\alpha_{\vec{k}} * \hat{\mathscr{G}}_{\vec{k}}
\right](\tau)
\left[
\hat{\Gamma}^\beta_{\vec{k}} * \hat{\mathscr{G}}_{\vec{k}}
\right](-\tau)
-(\alpha\leftrightarrow\beta)
\right\}
\end{aligned}
\end{equation*}
where $\left[ \hat{\Gamma}^\alpha_\vec{k} * \hat{\mathscr{G}}_\vec{k} \right](\tau)$ indicates convolution over imaginary times of the total vertex $\hat{\boldsymbol{\Gamma}}_\vec{k} = \vec{j}_{{\vec{k}}} + \delta\hat{\vec{j}}_{\vec{k}}(\tau)$ and the Green's function $\hat{\mathscr{G}}_\vec{k}$ dressed by spin fluctuations. The latter can be expanded in terms of the self-energy through a von Neumann series

\begin{equation*}
\begin{split}
        \hat{\mathscr{G}}_\vec{k}(\imath\omega_n) &= \hat{G}_\vec{k}(\imath\omega_n) \sum_{m=0}^\infty \left[ \hat{\Sigma}^s_\vec{k}(\imath\omega_n) \hat{G}_\vec{k}(\imath\omega_n) \right]^m \\ &= \hat{G}_\vec{k}(\imath\omega_n) + \hat{G}_\vec{k}^{(1)}(\imath\omega_n) + \hat{G}_\vec{k}^{(2)}(\imath\omega_n) + \cdots.
    \end{split}
\end{equation*}

\begin{figure*}[t]
    \centering
    \includegraphics[width=0.7\textwidth]{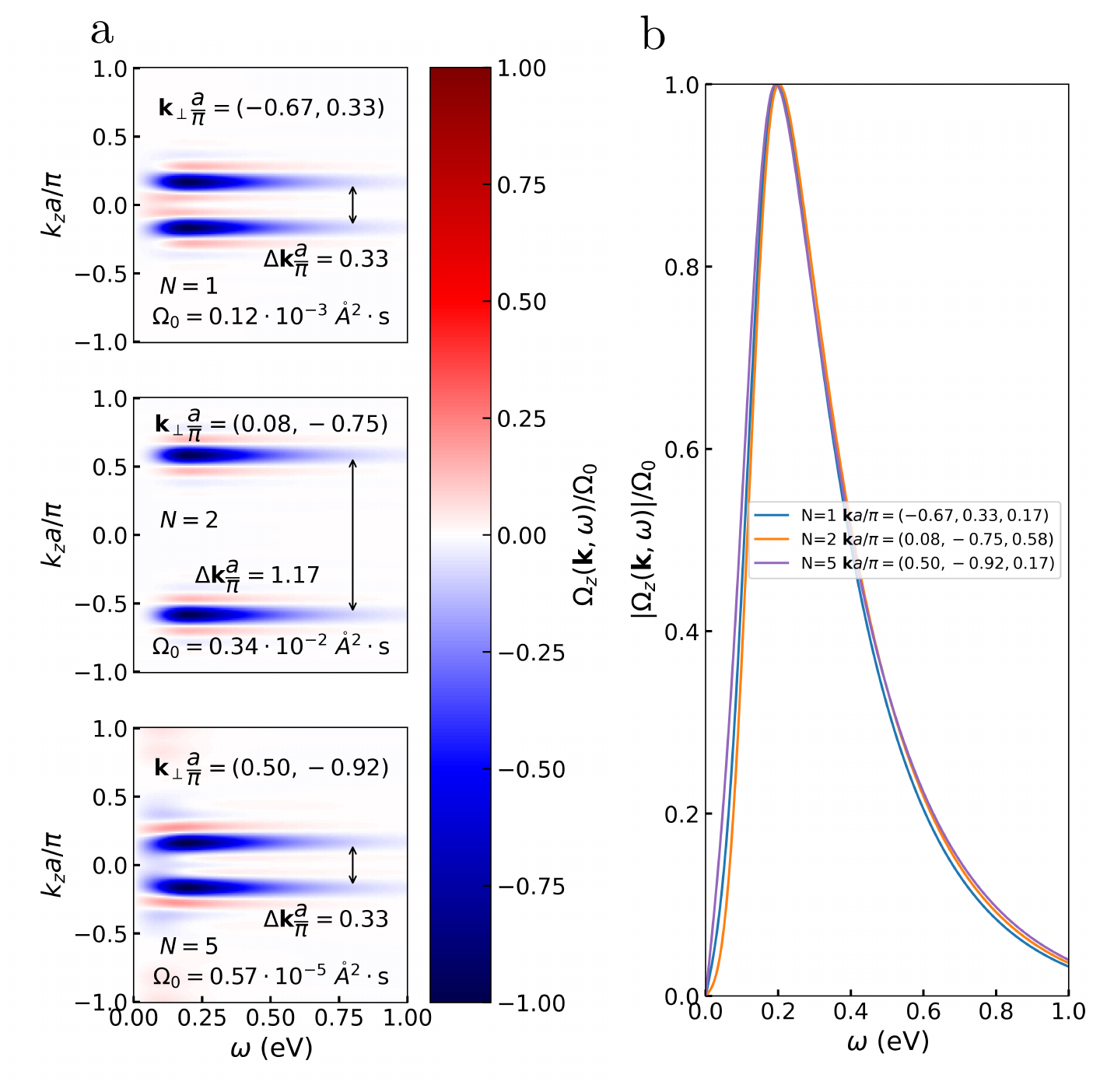}
    \caption{
(a) The spectral curvature, defined by $\Omega_z(\vec{k},\omega) = \dfrac{\Im\left[ \boldsymbol{\sigma}^{as}(\vec{k},\omega) \right]}{\omega}$ (Equation \ref{eq:geomcurv}) is shown in momentum-frequency space for three different occupancies of the $t_2g$ manifold, $N = 1, 2$ and 5. (b) Profiles of the geometric curvature showing its frequency dependence for different orbital occupancies, at specific momenta (indicated in the legend) where the curvature is maximum.     
}
    \label{fig:spectralberry}
\end{figure*}

It can be shown that the leading nonvanishing contribution appears at second order in the dressed vertex correction (Appendix \ref{append_lderivationantisymsuscept}). Defining the dressed self-energy propagator as
\begin{equation} \label{eq:dressed_F}
\hat{\mathscr{F}}_{\vec{k}}(\imath\omega_n)
=
\hat{G}_{\vec{k}}(\imath\omega_n)\,
\hat{\Sigma}^{s}_{\vec{k}}(\imath\omega_n)\,
\hat{G}_{\vec{k}}(\imath\omega_n)
\end{equation}
the antisymmetric susceptibility is expressed as

\begin{equation}\label{eq:xas}
\begin{aligned}
\boldsymbol{\chi}^{as}(\vec{k},\tau)
&= -\mathrm{tr} \left\{ \left[ \delta\hat{\vec{j}}_\vec{k} * \hat{G}_\vec{k} \right](\tau) \right.
\\
&\phantom{=}\hspace{24pt}
\left. \times \left[ \delta\hat{\vec{j}}_\vec{k} * \hat{G}_\vec{k} \right](-\tau) \right\}
\\
&\phantom{=}
-\vec{j}_{\vec{k}} \times
\mathrm{tr}\!\Big\{
\hat{\mathscr{F}}_{\vec{k}}(\tau)
\big[\delta\hat{\vec{j}}_{\vec{k}} * \hat{\tilde{G}}_{\vec{k}}\big](-\tau)
\\
&\phantom{=}\hspace{42pt}
- \hat{\mathscr{F}}_{\vec{k}}(-\tau)
\big[\delta\hat{\vec{j}}_{\vec{k}} * \hat{\tilde{G}}_{\vec{k}}\big](\tau)
\\
&\phantom{=}\hspace{42pt}
+ \hat{\tilde{G}}_{\vec{k}}(\tau)
\big[\delta\hat{\vec{j}}_{\vec{k}} * \hat{\mathscr{F}}_{\vec{k}}\big](-\tau)
\\
&\phantom{=}\hspace{42pt}
- \hat{\tilde{G}}_{\vec{k}}(-\tau)
\big[\delta\hat{\vec{j}}_{\vec{k}} * \hat{\mathscr{F}}_{\vec{k}}\big](\tau)
\Big\}.
\end{aligned}
\end{equation}
which can be continued analytically to real frequencies $\boldsymbol{\chi}^{as}(\vec{k},\omega)$. Crucially, due to the retardation effects introduced by fluctuations, Equation \ref{eq:xas} includes convolution terms of the type $\left( \delta \hat{\vec{j}}_{\mathbf{k}} * X_{\mathbf{k}} \right)(\tau)=\int d\tau'\;\delta \hat{\vec{j}}_{\mathbf{k}}(\tau - \tau')\,X_{\mathbf{k}}(\tau')$, where $X_{\mathbf{k}}=\hat{\mathscr{F}}_\vec{k}(\tau)$ or  $X_{\mathbf{k}}=\hat{G}_{\vec{k}}(\tau)$ (Appendix \ref{append_nonlocaltimecorrels}). Therefore, the cyclical properties of the trace cannot be used to map the $\mathrm{tr}\Big\{\hat{\mathscr{F}}_\vec{k}(\tau)\!\left[ \delta\hat{\vec{j}}_\vec{k} * \hat{G}_\vec{k} \right](-\tau)\!\Big\}$ onto the $\mathrm{tr}\Big\{
\hat{\mathscr{G}}_\vec{k}(\tau)\!\left[ \delta\hat{\vec{j}}_\vec{k} * \hat{\mathscr{F}}_\vec{k} \right](-\tau)\!\Big\}$ terms (which would make the structure traceless). In contrast, when vertices are instantaneous (as in Equation \ref{eq:bare}), the cyclical properties of the trace can be used to map the two fermionic lines to each other in the antisymmetrized bubble, making the corresponding algebraic structure traceless. 

These observations show that non-local time correlations are an essential ingredient to allow non-zero antisymmetric susceptibility. Additionally, as mentioned previously, transverse quantum fluctuations are also a crucial ingredient. The latter can be intuitively understood by observing that transverse spin ladder operators $S_+$, $S_-$ are necessary to give a nontrivial structure to spin-orbit coupling, otherwise its matrix reduces to the diagonal sector coming from $L_zS_z$ products, for which propagators and vertices, even if dressed by interactions, can be represented in the same spin-orbital basis, suppressing any relevant commutators. Therefore, both non-commutative transverse quantum fluctuations as well as time-non-locality are necessary ingredients to obtain nonzero spectral weight for the antisymmetric susceptibility.

\section{ANTISYMMETRIC SUSCEPTIBILITY AND DYNAMICAL BERRY CURVATURE} \label{sec:Berry_curv}
\subsection{Spectral and geometric Berry curvature}
    Henceforth, we explore the connection between the antisymmetric susceptibility channel and the geometric curvature, establishing a rigorous connection with the geometry in the response function. For that, we distinguish between the spectral Berry curvature and the dressed geometric curvature, the latter being the geometric object of reference. Since the emergence of spectral weight $\boldsymbol{\chi}^{as}(\vec{k},\omega)$ is driven by fluctuations, the associated curvature must be described with geometric models in which interaction effects are included \cite{souza2000polarization, ozawa2019probing, salerno2023drude, kashihara2023quantum, onishi2024quantum, chen2022measurement}. Among possible extensions of quantum geometry to interacting systems \cite{kashihara2023quantum, sukhachov2025effect}, the dressed quantum metric approach \cite{chen2022measurement} seems to provide good approximations to the quantum metric, at least for flat-band and weakly dispersive systems \cite{sukhachov2025effect}. We therefore follow the interaction-dressed quantum geometry, which relies on the charge polarization susceptibility to define self-energy corrections \cite{chen2022measurement}. Following this approach, a frequency-resolved spectral curvature can be defined from the antisymmetric component of the conductivity $\Omega^\gamma(\vec{k},\omega) = \dfrac{1}{\omega}\sum_{\alpha,\beta}\epsilon_{\alpha\beta\gamma}\Im \left[\sigma^{\alpha\beta}(\vec{k},\omega)\right]$. Its integration in frequency gives rise to a dressed curvature $\Omega^{\gamma}(\mathbf{k})=\int_{0}^{\infty} d\omega\,\Omega^{\gamma}(\mathbf{k},\omega)$, which reduces to the conventional Berry curvature in the noninteracting zero-temperature limit \cite{chen2022measurement}.

In view of these considerations, we therefore connect the spectral curvature $\Omega^{\gamma}(\mathbf{k},\omega)$ with the antisymmetric susceptibility $\boldsymbol{\chi}^{as}(\vec{k},\omega)$ defined in Equation \ref{eq:xas}. We can derive an expression (via analytic continuation) for the antisymmetric conductivity $\boldsymbol{\sigma}^{as}(\vec{k},\omega)$, whose real and imaginary parts are linked by Kramers-Kronig relations. This allows us to define an associated spectral Berry curvature
\begin{align}\label{eq:geomcurv}
    \boldsymbol{\Omega}(\vec{k},\omega) &= -\dfrac{\Im\left[ \boldsymbol{\sigma}^{as}(\vec{k},\omega) \right]}{\pi\omega}
\end{align}

In turn, the dressed geometric curvature can be obtained integrating in frequency
\begin{align}\label{eq:dressedberry}
    \begin{split}
    \boldsymbol{\Omega(\vec{k})} &= \int_0^\infty d\omega \ \boldsymbol{\Omega}(\vec{k},\omega) \\ &= -\int_0^\infty d\omega \ \dfrac{\Im\left[ \boldsymbol{\sigma}^{as}(\vec{k},\omega) \right]}{\pi\omega}.
    \end{split}
\end{align}

\subsection{Evaluation of spectral and geometric Berry curvatures}
The geometric curvature $\boldsymbol{\Omega(\vec{k})}$ generated by fluctuations requires reducing the symmetry below cubic, since otherwise it vanishes under $O_h$ point symmetry (Appendix \ref{append_symconstraints}). To evaluate its magnitude, we use a minimal model that assumes a symmetry-lowering process from a parent cubic lattice mediated by Jahn-Teller modes. Note that, while the Berry curvature generated by this mechanism is expected to be relatively weak in 4$d$-5$d$ systems due to the strong reduction of Jahn-Teller distortions under strong spin-orbit coupling  \cite{streltsov2020jahn, minarro2025emergent}, it nevertheless represents a lower bound for its value. Real materials with likely larger anisotropies in the hopping amplitudes and lower symmetries may eventually display curvatures that are stronger.

We therefore consider a cubic lattice where the symmetry is reduced by the stabilization of Jahn-Teller modes, whose energy is found using the Galiztkii-Migdal model (Appendix \ref{append_symconstraints}). The energy of these modes is typically a few meV up to around 10 meV \cite{minarro2025emergent}, whereas the corresponding anisotropy of the hopping amplitudes is $\delta t^\alpha/t \sim 0.1 - 1 \%$. Figure \ref{fig:berrymaps} shows the maps of the geometric curvature $\Omega_z(\vec{k})$ calculated in the $k_x - k_y$ plane perpendicular to the $\hat{z}$ axis of the tetragonal Jahn-Teller distortions for different orbital occupancies. Figure \ref{fig:berrymaps}a shows $\Omega_z(\vec{k})$ mapped in the $k_z=0$ (left column) and $k_z=\frac{\pi}{a}$ planes (right column), with high-symmetry points of the Brillouin zones indicated in the maps. On the other hand, Figure \ref{fig:berrymaps}b shows the map of $\Omega_z(\vec{k})$ plotted in the $k_z$ planes where the curvature reaches its maximum.

Two main observations can be drawn from inspecting these Figures. First, the finite curvature is concentrated in localized regions in momentum space, while it vanishes elsewhere. This arises from the $D_{4h}$ point symmetry imposed by Jahn-Teller distortions, which forces the dynamic curvature to vanish in specific regions of the Brillouin zone. For instance, diagonal and vertical mirror symmetries force the antisymmetric component $\Omega_z(\vec{k}) \propto \chi^{\mathrm{as}}_{xy}(\vec{k})$ to vanish in the planes $k_x=0,\pi$, $k_y=0,\pi$ and $k_x=\pm k_y$, as well as in points of high-symmetry such as $\Gamma, M, X$ or $R$. On the other hand, Figure \ref{fig:berryprojections} shows the projections of $\Omega_z(\vec{k})$ along different orientations perpendicular to different planes, confirming its confinement in all orientations in momentum space. 

On the other hand, the peak values of the curvature (around $10^{3}-10^{6}$ \AA$^2$, see Figure \ref{fig:berrymaps}b) are in a similar range to those reported near avoided crossings or nearly degenerate bands in other systems, including topological semimetals or moiré systems \cite{sheoran2023manipulation, bhowal2019electronic, kluczyk2024coexistence, xuan2020valley, zhang2022giant}. As discussed in the next section, this emergent curvature can be unambiguously detected by inelastic spectroscopies. Candidate materials to observe genuine fluctuation-driven curvatures may include centrosymmetric iridates \cite{foyevtsova2013ab} and other spin-orbit coupled systems such as $\alpha$-RuCl \cite{winter2016challenges}, which display bond-dependent hopping anisotropies well above a few per cent, resulting from octahedral rotations and bond-angle distortions, with a diversity of centrosymmetric structures \cite{di2016magnetic, hogan2016structural, stahl2024pressure}, which could offer a platform to probe selectively the dynamical curvature (provided that we probe the systems in their nonmagnetic or paramagnetic state).

In Figure \ref{fig:spectralberry} we additionally plot the spectral Berry curvature $\Omega_z(\vec{k}, \omega)$ for different orbital occupations along the path $k_z \in \left[-\frac{\pi}{a}, +\frac{\pi}{a}\right]$ in the Brillouin zone. Figure \ref{fig:spectralberry}a maps the peak value of $\Omega_z(\vec{k}, \omega)$, which, depending on orbital occupancy, is obtained at different planes $\frac{k_\perp a}{\pi}$ perpendicular to $k_z$, showing two localized areas where the spectral curvature is concentrated at low frequencies. Figure \ref{fig:spectralberry}b shows the frequency dependence on the spectral curvature $\Omega_z(\vec{k}, \omega)$ normalized to its peak value. The observed frequency dependence is consistent with a spectral curvature generated by coupling to spin fluctuations, whose energy is comparable to the width of the spectral weights of $\Omega_z(\vec{k}, \omega)$ shown in \ref{fig:spectralberry}b.

\section{PREDICTED RIXS SIGNATURES OF DYNAMICAL CURVATURE}\label{sec:signatures_RIXS}

So far, we have shown that the spectral curvature generated by fluctuations emerges in localized regions of the momentum-frequency space of the antisymmetric linear optical susceptibility. However, as mentioned above, optical spectroscopies do not allow to probe this curvature due to the nearly vanishing momentum transfer $\vec{q} \approx 0$ of optical photons. Here we address the following question: can other probes, in particular, resonant inelastic x-ray scattering spectroscopy (RIXS), enable the direct detection of spectral curvature generated genuinely from coupling to collective fluctuations?

\begin{figure}[!tbh]
    \centering
    \includegraphics[width=0.8\columnwidth]{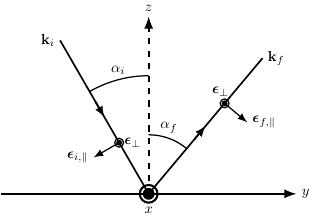}
    \caption{Schematic representation of the configuration of resonant inelastic x-ray scattering (RIXS) experiment. Parameters $\alpha_i, \alpha_f$ represent angles of incidence of incoming and and outgoing beams, $\vec{k}_i, \vec{k}_f$ their propagation vectors, and $\boldsymbol{\epsilon}_{\parallel}, \boldsymbol{\epsilon}_{\perp}$ the polarization vectors (note that $\boldsymbol{\epsilon}_{i,\perp}=\boldsymbol{\epsilon}_{f,\perp}=\boldsymbol{\epsilon}_{\perp}$). The scattering plane is $xz$.}
    \label{fig:rixs-setup}
\end{figure}

To address this question, we consider a RIXS direct scattering process in the ultrafast collision approximation, used in heavy transition metal compounds where the intermediate core-hole state lifetimes may be significantly shorter than the low-energy excitations of interest \cite{ament2011resonant, braicovich2020determining, jeong2017direct}. 
In this approximation, the Kramers-Heisenberg resolvent can be approximated to $[\omega_{\rm in}-(E_n-E_g)+i\Gamma]^{-1}\approx {i\Gamma}^{-1}$, because the core-hole state lifetime $\Gamma^{-1}$ is much larger than the energies of the incoming photons ($\omega_{\rm in}$), ground state ($E_g$) and intermediate states ($E_n$).  Therefore, in this limit, the resolvent can be treated as a constant and absorbed into an effective RIXS scattering vertex. Consequently, we define a scattering correlation function involving momentum transfer $\vec{q}$ as
\begin{align} \label{eq:ica-corr-qmain}
    \begin{split}
    \Pi_\vec{q}(\tau) &= -\dfrac{1}{\mathcal{N}} \sum_\vec{k} \mathrm{tr}\left[ \hat{\Lambda}\hat{\mathscr{G}}_\vec{k}(\tau) \hat{\Lambda}^\dagger \hat{\mathscr{G}}_{\vec{k}+\vec{q}}(-\tau) \right] 
    .
    \end{split}
\end{align}
where $\hat{\mathscr{G}}_\vec{k}(\tau)$ are Green's functions dressed by spin fluctuations, and the transition channel $\hat{\Lambda}_i = \hat{D}_i \hat{D}'{}_i^\dagger$ is defined through dipole operators $\hat{D}_i = \boldsymbol{\epsilon} \cdot \hat{\vec{r}}_i$, where $\boldsymbol{\epsilon}$ is the polarization vector, and $\hat{\vec{r}}_i$ the local position operator. 

We consider a scattering process in the $x-z$ plane (Figure \ref{fig:rixs-setup}), where the propagating wavevectors of the incoming and outgoing beams satisfy the condition $|\vec{k}_i| \approx |\vec{k}_f| \equiv k$. The allowed momentum transfer is given by $\vec{q} = \vec{k}_i - \vec{k}_f = -2k\sin\gamma\left[ \cos\delta \vec{y} - \sin\delta \vec{z} \right]$, where $2\gamma = \alpha_i + \alpha_f$ and $2\delta = \alpha_i - \alpha_f = 2\alpha_i-2\gamma$, and $\alpha_i , \alpha_f$ are, respectively, the angles of incidence and reflection. Additionally, the beam polarizations can be encoded using two angles $\theta$ and $\phi$ in the Bloch sphere, so that the polarization is expressed as $\boldsymbol{\epsilon}(\theta,\phi,\alpha) = \cos\dfrac{\theta}{2}\boldsymbol{\epsilon}_\perp + \mathrm{e}^{\imath\phi}\sin\dfrac{\theta}{2}\boldsymbol{\epsilon}_\parallel(\alpha)$, where $\boldsymbol{\epsilon}_\perp = \vec{x}$, $\boldsymbol{\epsilon}_{i,\parallel}(\alpha_i) = -\cos\alpha_i \vec{y} - \sin\alpha_i \vec{z}$ and $\boldsymbol{\epsilon}_{f,\parallel}(\alpha_f) = \cos\alpha_f \vec{y} - \sin\alpha_f \vec{z}$. 

To extract the antisymmetric signal, the polarization of the reflected light must be perpendicular to that of the incident beam. Consequently, after defining $\theta\equiv\theta_i = \pi-
\theta_f$ and $\phi\equiv\phi_i=\phi_f-\pi$, we can compute the antisymmetric component of the scattering function as
\begin{align}
\begin{split}
\Pi^{\mathrm{as}}_{\vec q}(\theta,\phi,\gamma,\delta;\tau)
&= \Pi_{\vec q}(\theta,\phi,\gamma,\delta;\tau) \\
&\quad - \Pi_{\vec q}(\pi-\theta,\phi+\pi,\gamma,\delta;\tau).
\end{split}
\end{align}

In general, this antisymmetric response can be nonzero even if spin fluctuations are ignored. The reason is that the experimental (scattering) geometry (e.g., plane of incidence and polarization configurations) can generally contribute to the antisymmetric channel even if the system is $\mathcal{P}-\mathcal{T}$ symmetric. It is therefore important to reduce as much as possible any residual contribution not directly related to the curvature induced by fluctuations. Interestingly, a condition can be found in which the polarization angles satisfy $\theta_i=\theta_f=\pi/2$, which nullifies the antisymmetric channel (ignoring fluctuations) regardless of the angles of incidence and reflection (Appendix \ref{append_polarizationasymchannel}). In this particular condition, the polarizations of the incident and emerging beams are described by $\boldsymbol{\epsilon}(\theta=\pi/2,\phi,\alpha) = \dfrac{\sqrt{2}}{2}[\boldsymbol{\epsilon}_\perp + \mathrm{e}^{\imath\phi}\boldsymbol{\epsilon}_\parallel(\alpha)]$. Consequently, we can suppress any unwanted contribution from the geometric scattering settings by choosing the polarizations $\boldsymbol{\epsilon}(\theta=\pi/2,0,\alpha_i)$ and $\boldsymbol{\epsilon}(\theta=\pi/2,\pi,\alpha_f)$ in $\pm45^\circ$ with respect to the scattering plane. 

Having established the optimal parameters of the scattering geometry, we now find a mathematical expression for the antisymmetric scattering function. For that, we focus on $L$-edge transitions, where core electrons in $p$-orbitals jump to valence band  $t_{2g}$ states. The corresponding dipole matrix elements can be computed as

\begin{align}
\langle d_{\mu\nu} \mid \hat{r}_\alpha \mid p_\kappa\rangle
&= \dfrac{3\sqrt{5}}{4\pi}
\int_0^\infty dr\, r^2 R_{n2}(r)R_{n1}(r)
\nonumber\\
&\quad\times
\int_{4\pi} d\Omega\,
\dfrac{r_\mu r_\nu r_\alpha r_\kappa}{r^4}
\nonumber\\
&= \dfrac{N_n}{\sqrt{5}}
\left[
\delta_{\mu\alpha}\delta_{\nu\kappa}
+
\delta_{\mu\kappa}\delta_{\nu\alpha}
\right].
\end{align}
where factor $N_n$ is the radial integral depending on quantum number $n$. With this result, the expression of the matrix for the dipole operator expressed in the $t_{2g}$ basis $\left\{ \lvert yz\rangle,\ \lvert zx\rangle,\ \lvert xy\rangle \right\}$ is

\begin{align*}
    \hat{D}_i &= \dfrac{N_n}{\sqrt{5}} \begin{pmatrix}
        0 & \epsilon_z & \epsilon_y \\
        \epsilon_z & 0 & \epsilon_x \\
        \epsilon_y & \epsilon_x & 0
    \end{pmatrix}
\end{align*}

where, to simplify notation, the Cartesian components of the polarization are written without the explicit dependence on Bloch sphere or scattering angles. Then, the expression for the transition channel operator is written as

\begin{align*}
    \hat{\Lambda}_i &= \dfrac{N_n^2}{5} \sum_{\mu,\nu} \left[ \boldsymbol{\epsilon}_f^* \cdot \boldsymbol{\epsilon}_i \delta_{\mu\nu} + (-1)^{\delta_{\mu\nu}} \epsilon_{f,\mu}^* \epsilon_{i,\nu} \right] |\mu_i\rangle\langle\nu_i|.
\end{align*}

where $|\mu_i\rangle\langle\nu_i|$ are orbital transition operators written in the $t_{2g}$ basis. Now, using the values of the polarization angles that suppress the background not related to fluctuations, we can reformulate the expressions for the transition channel operators as follows

\begin{figure*}[t]
    \centering
    \includegraphics[width=0.92\textwidth]{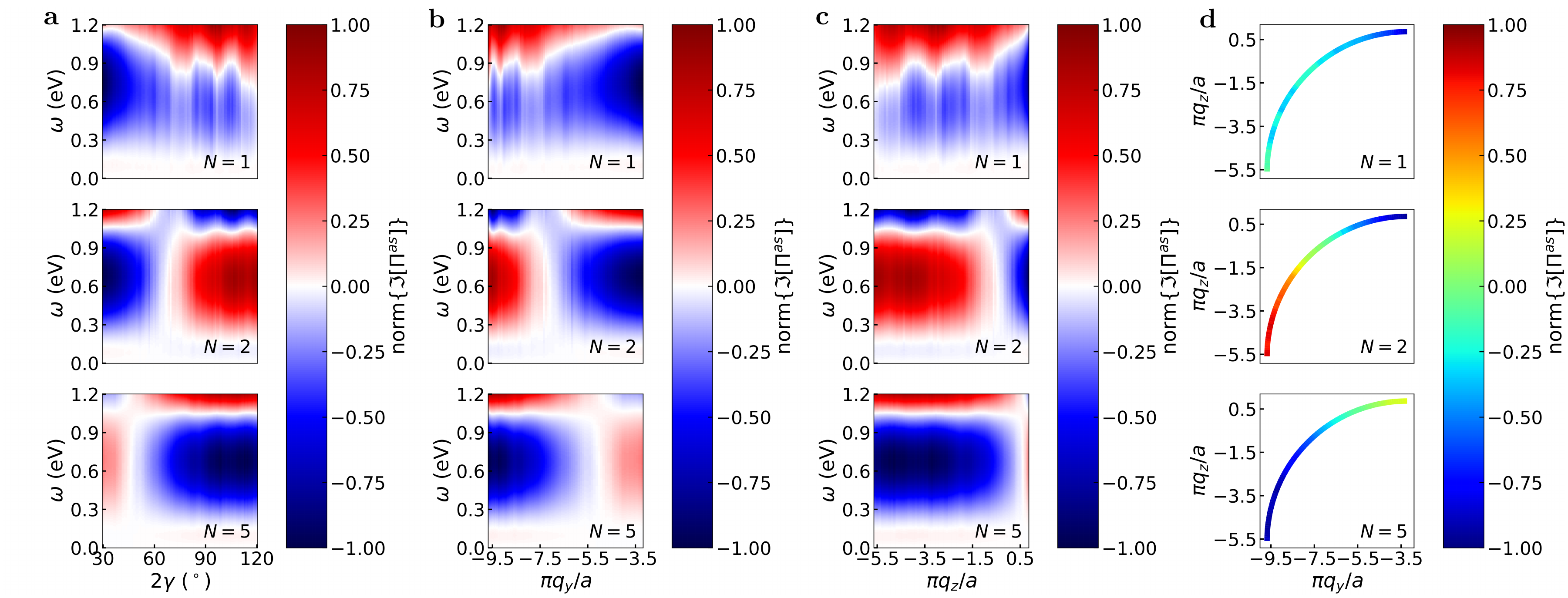}
    \caption{
(a) The antisymmetric scattering function $\Pi_{\vec{q}}^{\mathrm{as},(1)}(\gamma,\delta;\tau)$ (Equation \ref{eq:scattering_funct_with_fluctuations}) is mapped as a function of frequency $\omega$ and scattering angle $2\gamma = \alpha_i + \alpha_f$, where $\alpha_i , \alpha_f$ are, respectively, the angles of incidence and reflection. This function corresponds to scattering parameters chosen so that $\Pi_{\vec{q}}^{\mathrm{as},(1)}(\gamma,\delta;\tau)$ is only sensitive to contributions from fluctuations. In the calculations, the angle of incidence was fixed at $\alpha_i = 30^\circ$, so the explored range of $2\gamma$ implies scanning the angle of reflectance in the interval $0^\circ\leq \alpha_f\leq90^\circ$. The values of $\Pi_{\vec{q}}^{\mathrm{as},(1)}(\gamma,\delta;\tau)$ are normalized to the maximum value. Panels (b) and (c) show the same data plotted as a function of scanned momentum transfers $q_y$ and $q_z$. A lattice parameter $a = 4\mathring{\mathrm{A}}$ is assumed. (d) Trajectory of the momentum transfer in the $q_y-q_z$ plane scanned at $\omega=0.6$ eV, corresponding to the maps shown in panels (b)-(c) .
}
    \label{fig:rixsmaps}
\end{figure*}

\begin{widetext} 
\begin{align} \label{eq:lambda_plus_minus}
\hat{\Lambda}_\pm(\gamma,\delta)
&= \dfrac{N_n^2}{10}
\begin{pmatrix}
-\cos(2\gamma) & \mp\cos(\gamma+\delta) & \mp\sin(\gamma+\delta) \\
\pm\cos(\gamma-\delta) &
1 - \frac{1}{2}[3\cos(2\gamma)+\cos(2\delta)] &
\frac{1}{2}[\sin(2\gamma)+\sin(2\delta)] \\
\mp\sin(\gamma-\delta) &
-\frac{1}{2}[\sin(2\gamma)-\sin(2\delta)] &
1 - \frac{1}{2}[3\cos(2\gamma)-\cos(2\delta)]
\end{pmatrix}
\nonumber\\
&= \dfrac{N_n^2}{10} \Big[
\pm \sin\gamma \sin\delta\,\hat{\lambda}^1
\mp \imath \cos\gamma \cos\delta\,\hat{\lambda}^2
\mp \sin\gamma \cos\delta\,\hat{\lambda}^4
\mp \imath \cos\gamma \sin\delta\,\hat{\lambda}^5
\nonumber\\
&\hspace{2.8cm}
+ \dfrac{1}{2}\sin(2\delta)\,\hat{\lambda}^6
+ \dfrac{\imath}{2}\sin(2\gamma)\,\hat{\lambda}^7
+ \dfrac{2-4\cos(2\gamma)}{3}\,\hat{\lambda}^0
\nonumber\\
&\hspace{2.8cm}
+ \dfrac{\cos(2\gamma)+\cos(2\delta)-2}{4}\,\hat{\lambda}^3
+ \dfrac{\cos(2\gamma)-3\cos(2\delta)-2}{4\sqrt{3}}\,\hat{\lambda}^8
\Big]
\nonumber\\
&= \dfrac{N_n^2}{10}\sum_a c_\pm^a(\gamma,\delta)\,\hat{\lambda}^a.
\end{align}
\end{widetext}

In this expression, the operator $\hat{\Lambda}_+(\gamma,\delta)$ describes the experiment in which the incident light is linearly polarized at $+45^\circ$ with respect to the scattering plane and the reflected light is polarized at $-45^\circ$, i.e. $\phi_i=0$ and $\phi_f=\pi$. On the other hand, the operator $\hat{\Lambda}_-(\gamma,\delta)$ describes the reverse situation, namely, the incident beam is polarized at $-45^\circ$ and the reflected beam is polarized at $+45^\circ$ with respect to the scattering plane, i.e. $\phi_i=\pi$ and $\phi_f = 0$. In Equation \ref{eq:lambda_plus_minus} $\hat{\lambda}^{a}$ are Gell-Mann matrices, while $c_\pm^a(\gamma,\delta)$ are appropriate combinations of trigonometric functions. 

Using these polarization settings, the antisymmetric scattering function can be expanded to the first order of the spin projections, giving as a result

\begin{widetext}
\begin{align}
\Pi_{\vec{q}}^{\mathrm{as},(1)}(\gamma,\delta;\tau)
&= -\dfrac{1}{\mathcal{N}}
\sum_{\vec{k},\,s=\{+1,-1\}}
(-1)^{\frac{s+1}{2}}
\mathrm{tr}\Big[
\hat{\Lambda}_s(\gamma,\delta)\hat{\tilde{G}}_{\vec{k}}(\tau)
\hat{\Lambda}_s^\dagger(\gamma,\delta)
\hat{\mathscr{F}}_{\vec{k}+\vec{q}(\gamma,\delta)}(-\tau)
\nonumber\\
&\hspace{3.2cm}
+\hat{\Lambda}_s(\gamma,\delta)\hat{\mathscr{F}}_{\vec{k}}(\tau)
\hat{\Lambda}_s^\dagger(\gamma,\delta)
\hat{\tilde{G}}_{\vec{k}+\vec{q}(\gamma,\delta)}(-\tau)
\Big]
\nonumber\\
&=
\dfrac{N_n^4}{50}
\sum_{a,b,\,s=\{+1,-1\}}
(-1)^{\frac{s+1}{2}}
\Big[
c_s^a(\gamma,\delta)c_s^b(\gamma,\delta)^*
\,\Xi^{ab}(\gamma,\delta;\tau)
\nonumber\\
&\hspace{3.2cm}
+c_s^a(\gamma,\delta)^*c_s^b(\gamma,\delta)
\,\Xi^{ab}(\gamma,\delta;-\tau)
\Big].
\end{align}
\end{widetext}
where we define the function $\Xi_{\vec{q}}^{ab}(\gamma,\delta;\tau) = -\dfrac{1}{\mathcal{N}} \sum_\vec{k} \mathrm{tr}\left[ \hat{\lambda}^a \hat{\tilde{G}}_\vec{k}(\tau) \hat{\lambda}^b \hat{\mathscr{F}}_{\vec{k}+\vec{q}(\gamma,\delta)}(-\tau) \right]$, and make use of the dressed self-energy propagator $\hat{\mathscr{F}}_{\vec{k}+\vec{q}(\gamma,\delta)}(-\tau)$ defined in Equation \ref{eq:dressed_F}. Finally, after defining the matrix $\hat{\mathcal{C}}(\gamma,\delta)$ with elements $\mathcal{C}_{ba}(\gamma,\delta) = c_+^a(\gamma,\delta)c_+^b(\gamma,\delta)^* - c_-^a(\gamma,\delta)c_-^b(\gamma,\delta)^*$ we obtain the following expression for the antisymmetric scattering function

\begin{align} \label{eq:scattering_funct_with_fluctuations}
\Pi_{\vec{q}}^{\mathrm{as},(1)}(\gamma,\delta;\tau)
&= \dfrac{N_n^4}{50}
\mathrm{tr}\Big[
\hat{\mathcal{C}}(\gamma,\delta)\hat{\Xi}_{\vec{q}}(\gamma,\delta;\tau)
\nonumber\\
&\hspace{1.7cm}
+ \hat{\mathcal{C}}^\dagger(\gamma,\delta)\hat{\Xi}_{\vec{q}}(\gamma,\delta;-\tau)
\Big].
\end{align}

We then arrive at an expression for the antisymmetric scattering function $\Pi_{\vec{q}}^{\mathrm{as},(1)}(\gamma,\delta;\tau)$, which depends on the momentum transfer $\vec{q}$, imaginary time $\tau$ and scattering angles $\gamma, \delta$, and is solely sensitive to those excitations that arise genuinely from the coupling to spin fluctuations, filtering out any other excitations of different origin.

\begin{figure}[t]
    \centering
    \includegraphics[width=\columnwidth]{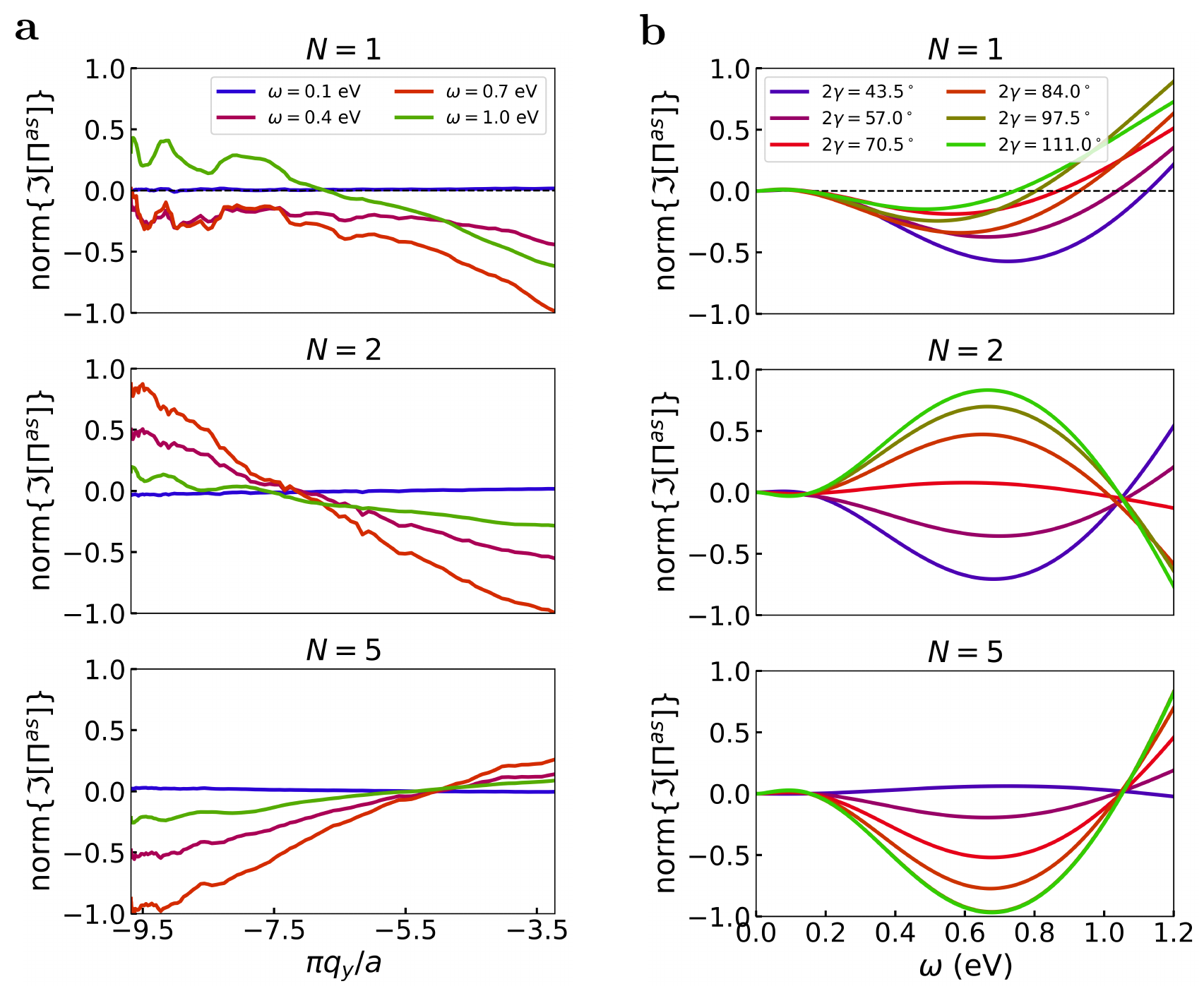}
    \caption{
    (a) Profiles of the scattering function $\Pi_{\vec{q}}^{\mathrm{as},(1)}(\gamma,\delta;\tau)$ defined in Equation \ref{eq:scattering_funct_with_fluctuations}, obtained at different energy cuts indicated in the legend. Their dependence on the momentum transfer $q_y$ is shown for different orbital occupations. The data are normalized to the maximum value. (b) Profiles of $\Pi_{\vec{q}}^{\mathrm{as},(1)}(\gamma,\delta;\tau)$ displayed as a function of energy for fixed values of the momentum transfer $q_y$ indicated in the legend.  
    }
    \label{fig:rixsprofiles}
\end{figure}

Figure \ref{fig:rixsmaps}a shows the antisymmetric scattering response $\Pi_{\vec{q}}^{\mathrm{as},(1)}(\gamma,\delta;\tau)$ mapped for different orbital occupations as a function of frequency and scattering angle in the range $30^\circ \leq 2\gamma \leq 120^\circ$. In the calculations, the angle of incidence was fixed at $\alpha_i=30^\circ$, corresponding to a variation of the angle of the reflected beam in the interval $0^\circ \leq \alpha_f \leq 90^\circ$. RIXS maps are also shown in Figures \ref{fig:rixsmaps}b and \ref{fig:rixsmaps}c as a function of the scanned momentum transfers along $q_y$ and $q_z$, while the corresponding trajectory in the $q_y-q_z$ plane is shown in Figure \ref{fig:rixsmaps}d. Bearing in mind that our model is based on an ultrafast collision approximation, the RIXS signal is plotted for frequencies well below the inverse lifetime broadening determined from RIXS experiments, which is around $\Gamma \approx 3-5$ eV for typical L-edge transitions \cite{kim2018probing,igarashi2014analysis, jay2026metal}. Consequently, the maps in Figure \ref{fig:rixsmaps} display $\Pi_{\vec{q}}^{\mathrm{as},(1)}(\gamma,\delta;\tau)$ sufficiently below this frequency cutoff.

Some observations can be made with respect to these results. First, the data in Figure \ref{fig:rixsmaps} reflect the emergence of a finite antisymmetric RIXS signal, which would otherwise vanish in a $\mathcal{P}-\mathcal{T}$ system. By construction, the scattering function $\Pi_{\vec{q}}^{\mathrm{as},(1)}(\gamma,\delta;\tau)$ only probes the excitations that arise purely from the coupling to fluctuations. Interestingly, the most prominent observed RIXS signals are spread around energies $\approx 0.4 - 0.9$ eV (see Figure \ref{fig:rixsprofiles}), which is consistent with the typical range in which spin-orbital excitations are observed in heavy transition metal compounds \cite{kim2014excitonic, das2018spin}. This suggests that in our approximation, in which we consider the coupling to spin fluctuations, spin-orbit coupling is strong enough to drive spin-orbital excitations in the observed range. This is an important observation, since the observed antisymmetric spectral weight arises only through fluctuation-dressed dynamics, which would otherwise vanish if fluctuations were ignored. Finally, we observe that, contrary to optical spectroscopies with nearly vanishing momentum transfer, the selectivity introduced by finite-$\vec{q}$ inelastic spectroscopies allows directly to probe the geometric curvature generated dynamically by fluctuations through the antisymmetric RIXS scattering function $\Pi_{\vec{q}}^{\mathrm{as},(1)}(\gamma,\delta;\tau)$. 

\section{CONCLUSION}

In this work, we have shown that inelastic spectroscopies can be used to detect excitations related to the emergence of geometric curvature driven exclusively by dynamical coupling to fluctuations. A key ingredient is finite-$\vec{q}$ momentum transfer, which enables the detection of regions of finite spectral weight in momentum frequency space where fluctuation-driven geometries are generated. Thus, we show that manybody contributions to quantum geometry, which have recently received attention \cite{sukhachov2025effect, chen2022measurement, kashihara2023quantum, onishi2024quantum, guan2026exploring, jin2026experimental}, can be selectively detected by specific antisymmetric channels of the inelastic susceptibility. We further identify the non-commutative properties of transverse quantum fluctuations as well as non-local-time interactions as the generators of the effective gauge structure associated with the fluctuation-driven curvature. 

Our interpretation in terms of geometry is crucial: without it, we may lose an essential framework to understand some deep physical implications. For example, recent work emphasizes the relation between quantum Fisher information and quantum metric, which can be measured experimentally by resonant inelastic x-ray spectroscopy, providing information on the entanglement of electronic orbitals \cite{gao2025quantum, ren2024witnessing}. This is an example of how response functions allow access to quantum metric and quantum information \cite{gao2025quantum, yu2025quantum}. Our approach is complementary: rather than targeting contributions from the real part of the geometric tensor (the quantum metric), we focus on antisymmetric contributions that generate a dynamical curvature structure in the response functions, giving access to the imaginary part of the geometric tensor and the geometric phases driven by interactions with fluctuations. As a future perspective, one may also address how to detect manybody quantum contributions to the quantum metric, rather than curvature. Though the measurement of quantum metric is usually more challenging, its investigation may benefit from recent developments in angle-resolved photoemission spectroscopy \cite{kang2025measurements, jin2026experimental, kim2025direct} or ultrafast techniques \cite{verma2025framework}.

The work presented here should be viewed as a minimal theoretical approach. More quantitative models should rely on the inclusion of $e_g$ states \cite{paramekanti2018spin,stamokostas2018mixing, minarro2022prb, minarro2024spin} as well as contributions from charge transfer processes. Also, to explore these higher energy excitations, RIXS models should extend beyond the ultrafast collision approximation and include the effects of core-hole potentials \cite{ament2011resonant, kotani2001resonant}. In addition, material-specific contributions to the electronic structure should also be considered. Despite these limitations, the work described here shows that RIXS is a promising route to probe dynamical fluctuation-driven responses and associated geometric structures in correlated spin-orbit-coupled systems. \\

\setlength{\dbltextfloatsep}{8pt}  
\setlength{\dblfloatsep}{8pt}      

\textbf{ACKNOWLEDGMENTS}\\ This work was supported by Projects No. PID2023-152225NB-I00 and Severo Ochoa MATRANS42 (No. CEX2023-001263-S) of the Spanish Ministry of Science and Innovation (Grant No. MICIU/AEI/10.13039/501100011033 and FEDER, EU).

\vspace{0.5cm}

\textbf{DATA AVAILABILITY}\\
The data that support the findings of this article are not publicly available. The data are available from the authors upon reasonable request. All codes to reproduce, examine, and improve our proposed analysis are available at \url{https://doi.org/10.5281/zenodo.15527346} or from the corresponding authors upon request.

\appendix

\section{LATTICE MODEL}
\label{append_latticemodel}

\subsection{Computation of the optical susceptibility} \label{append_comput_opt_susc}
To compute the longitudinal and antisymmetric transverse part of the optical susceptibility, we work within the Kubo formalism. We focus on obtaining linear responses in inversion ($\mathcal{P}$) time-reversal ($\mathcal{T}$) symmetric systems. The susceptibility is determined from the retarded current-current correlation function. The optical conductivity tensor is defined as

\begin{equation}
\sigma_{ij}(\omega)=\frac{1}{i\omega}\chi_{ij}^{R}(\omega),
\end{equation}
where $\chi_{ij}^{R}(\omega)$ is the retarded current--current propagator. The longitudinal optical conductivity corresponds to the diagonal components of this tensor,
\begin{equation}
\sigma_{ii}(\omega)=\frac{1}{i\omega}\left[\chi_{ii}^{R}(\omega) - \dfrac{e^2\langle \hat{n} \rangle}{m}  \right],
\end{equation}

In addition to the longitudinal response, we consider the antisymmetric component of the optical susceptibility
\begin{equation}\label{eqchiantisym}
\begin{aligned}
\chi_{ij}^{as}(\omega)
&= \tfrac{1}{2}\big[\chi_{ij}(\omega)-\chi_{ji}(\omega)\big].
\end{aligned}
\end{equation}
This antisymmetric contribution encodes Hall-like optical responses and is directly connected to the spectral Berry curvature.

To evaluate these quantities, we first obtain the current--current correlator in imaginary time in the Matsubara branch
\begin{equation}
\chi_{\mu\nu}(\tau)
=
\left\langle
\hat{j}^{\mu}(\tau)\hat{j}^{\nu}(0)
\right\rangle .
\end{equation}
where $\hat{j}^{\mu}(\tau)$ is the $\mu$ component of the current operator in imaginary time. Within the diagrammatic expansion, the current-current correlator can be written as
\begin{equation}\label{eqcurrcurrcorrel}
\begin{aligned}
\chi_{\mu\nu}(\tau)
&=
\left\langle \hat{j}^{\mu} \right\rangle
\left\langle \hat{j}^{\nu} \right\rangle \\
&\quad
- \sum_{\vec{k}}
\mathrm{Tr}\!\left[
\hat{j}^{\mu}_{\vec{k}}
\hat{G}_{\vec{k}}(\tau)
\hat{j}^{\nu}_{\vec{k}}
\hat{G}_{\vec{k}}(-\tau)
\right].
\end{aligned}
\end{equation}

where, in equilibrium, the first term vanishes in the absence of external fields. Then, Fourier transforming to Matsubara frequencies and performing analytic continuation $i\omega_n\rightarrow\omega+i0^+$ yields the retarded propagator $\chi^R_{\mu\nu}(\omega)$, from which we obtain the optical conductivity and susceptibility as a function of real frequencies. In the following, we first describe how we obtain the propagators in imaginary time and, subsequently, how we continue analytically to obtain the propagators in real frequencies. 

\subsection{Lattice model in the Matsubara formalism} \label{append_matsubara}

To obtain the propagators in imaginary time, we define the Green's functions, which are solved self-consistently as follows
 
\[
[\hat{G}_{\vec{k}}(\tau)]_{\alpha\sigma,\alpha'\sigma'}
   = -\langle \mathcal{T}_\tau
     \hat{c}_{\vec{k}\alpha\sigma}(\tau)
     \hat{c}^\dagger_{\vec{k}\alpha'\sigma'}(0)
     \rangle.
\]
where $\alpha, \alpha'$ define orbitals in the $t_{2g}$ basis and $\sigma, \sigma'$ the spin degrees of freedom. To determine $\hat{G}_{\vec{k}}(\tau)$, we solve a Hamiltonian including hopping terms, spin--orbit coupling, and Kanamori electron-electron correlations, with the self-energy evaluated at the Hartree-Fock and second-Born levels (see Ref.~\cite{minarro2025emergent} for details). Motivated by recent observations that Jahn--Teller (JT) dynamics persist even in strongly spin-orbit-coupled systems~\cite{celiberti2024spin,vzivkovic2024dynamic,iwahara2023vibronic,iwahara2024dynamic,khaliullin2021exchange},
we also include $T_{2g}\otimes E_g$ vibronic interactions in the Hamiltonian. Thus, the Hamiltonian \cite{minarro2025emergent} that we analyze is:

\allowdisplaybreaks
\setlength{\jot}{2pt} 

\begin{align}
\hamil &= \hamil_{latt} + \hamil_{JT} + \hamil_{ph} + \hamil_{SO} + \hamil_{ee} \notag\\
\hamil_{latt} &= \sum_{i,j} \sum_{\mu,\nu} \sum_\sigma 
    t_{i\mu,j\nu} c_{i\alpha\sigma}^\dagger c_{j\beta\sigma} \notag\\
&= \sum_{\vec{k}} \sum_{\mu,\nu} \sum_\sigma 
    \varepsilon_{\mu\nu}(\vec{k}) c_{\vec{k}\alpha\sigma}^\dagger c_{\vec{k}\beta\sigma} \notag\\
\hamil_{ph} &= \omega_0 \sum_i \!\left( b_{i\phi}^\dagger b_{i\phi} + b_{i\theta}^\dagger b_{i\theta} \right) \notag\\
\hamil_{JT} &= \sum_{i,\eta \in \{3,8\},\alpha,\beta,\sigma,\sigma'}
    g_\eta \lambda^\eta_{\alpha\beta} c_{i\alpha\sigma}^\dagger c_{i\beta\sigma'}
    \left( b_{i\eta} + b_{i\eta}^\dagger \right) \notag\\
\hamil_{SO} &= \frac{\xi}{2} \sum_i \sum_{\sigma,\sigma'} \sum_{\mu,\nu \in d}
    \langle \mu|\vec{l}|\nu\rangle \!\cdot\! \langle \sigma|\boldsymbol{\sigma}|\sigma'\rangle\,
    c_{i\mu\sigma}^\dagger c_{i\nu\sigma'} \notag\\
\hamil_{ee} &= U \sum_{i,\alpha} n_{i\alpha\uparrow} n_{i\alpha\downarrow}
    + U' \sum_{i,\alpha \neq \gamma} n_{i\alpha\uparrow} n_{i\gamma\downarrow} \notag\\
&\quad + (U' - J) \sum_{i,\alpha < \gamma,\sigma} n_{i\alpha\sigma} n_{i\gamma\sigma} \notag\\
&\quad - J \sum_{i,\alpha \neq \gamma} c_{i\alpha\uparrow}^\dagger
    \left( c_{i\alpha\downarrow} c_{i\gamma\downarrow}^\dagger
    + c_{i\gamma\downarrow} c_{i\alpha\downarrow}^\dagger \right) c_{i\gamma\uparrow} \label{eq:H}
\end{align}

This model describes \( t_{2g} \) electrons in a lattice, where the kinetic energy is captured by \( \mathcal{H}_{\text{latt}} \), while \( \mathcal{H}_{\text{SO}} \) and \( \mathcal{H}_{\text{ee}} \) account for spin--orbit coupling and electron--electron interactions, respectively. Orbital indices are denoted by \( \mu, \nu \), spin indices by \( \sigma, \sigma' \), and \( i, j \) label lattice sites. The terms \( \mathcal{H}_{\text{ph}} \) and \( \mathcal{H}_{\text{JT}} \) describe, respectively, the energy of the Jahn--Teller active \( E_g \) phonon modes---characterized by frequency \( \omega_0 \)---and the coupling between \( t_{2g} \) orbitals and lattice distortions of orthorhombic (\( \eta = \phi \), \( Q_2 \)) and tetragonal (\( \eta = \theta \), \( Q_3 \)) symmetry, with coupling strengths \( g_\eta \)~\cite{streltsov2020jahn, bersuker2006jt, bersuker2013vibronic, khomskii2014}. The term \( \mathcal{H}_{\text{JT}} \) involves Gell--Mann matrices \( \lambda^\eta_{\alpha\beta} \), which encode the symmetry of the two \( t \otimes E \) Jahn--Teller modes, i.e., \( \eta = 3 \) corresponds to the orthorhombic (\( \phi \), \( Q_2 \)) mode, and \( \eta = 8 \) to the tetragonal (\( \theta \), \( Q_3 \)) mode.

To solve the Hamiltonian, we iterate the following self-consistent equations \cite{minarro2025emergent}:
\begin{align}
    \begin{split} \label{eq:GFself}
        \hat{G}(\imath\omega_n) &= \left[
        \hat{\mathfrak{G}}^{-1}(\imath\omega_n) - \left(\hat{\Sigma}^{\mathrm{HF}} 
        + \hat{\Sigma}^{(2)}(\imath\omega_n) \right. \right. \\
        &\phantom{=} \left. \left.
        + \hat{\Sigma}^{\mathrm{ep}}(\imath\omega_n) \right)
        \right]^{-1}
    \end{split} \\
    \label{eq:phself}
    D_\eta(\imath\nu_n) &= 
    \left[
    D^{(0)\,-1}_\eta(\imath\nu_n) - \Pi_\eta(\imath\nu_n)
    \right]^{-1}
\end{align}
Here, \( \hat{G} \) and \( \hat{\mathfrak{G}} \) are the dressed and bare Green's functions. The self-energy includes Hartree--Fock (\( \hat{\Sigma}^{\mathrm{HF}} \)), second-order Born approximation (\( \hat{\Sigma}^{(2)} \)) and electron--phonon (\( \hat{\Sigma}^{\mathrm{ep}} \)) terms. The phonon propagator \( D_\eta \) is renormalized by the self-energy \( \Pi_\eta \), and \( D^{(0)}_\eta \) denotes the bare propagator. Matsubara frequencies are \( \imath\omega_n \) (fermionic) and \( \imath\nu_n \) (bosonic).

\subsection{Analytic continuation} \label{append_analyticcontinuation}

To obtain the optical responses, we perform an analytic continuation from the imaginary to the real frequency domain. For that, we use the intermediate representation (IR) framework~\cite{chikano2018irbasis, shinaoka2021efficient, wallerberger2023sparseir}, using the fact that the susceptibility is related to the optical conductivity through a Fredholm integral equation of the first kind~\cite{jarrell1996bayesian, gull2011continuous},  
\[
\chi^p_{ij}(\tau)=-\frac{1}{\pi}\int_0^\infty d\omega\, K(\tau,\omega)\,\Re[\sigma^p_{ij}(\omega)],
\]

Here, $\chi^p_{ij}(\omega)$ and $\sigma^p_{ij}(\omega)$ are the paramagnetic susceptibility and conductivity, while the bosonic kernel is \(K(\tau,\omega)=\omega e^{-\tau\omega}/(1-e^{-\beta\omega})\). To solve this inverse problem, the kernel is compressed via singular value decomposition~\cite{chikano2018irbasis, shinaoka2021efficient, wallerberger2023sparseir}, 
\(K(\tau,\omega)=\sum_l U_l(\tau)S_lV_l(\omega)\), 
where \(U_l(\tau)\) and \(V_l(\omega)\) are orthonormal bases and \(S_l\) decay exponentially, so that only a few modes contribute significantly. In the IR basis, the integral equation becomes a linear system,  
\[
\chi^p_{ij;l} + \sum_m \mathcal{K}_{lm}\,\Xi_{ij;m}=0,
\]
where \(\chi^p_{ij;l}\) are the IR coefficients of the imaginary-time data, \(\mathcal{K}_{lm}\) are kernel matrix elements, and the spectral function is defined as \(\Xi_{ij}(\omega)=\Re[\sigma_{ij}(\omega)]/\pi\). The system is solved with Tikhonov regularization~\cite{ghanem2023tikhonov, schott2016continuation} to stabilize the inversion, and the full conductivity is then reconstructed, with its real and imaginary parts related by the Kramers--Kronig relation.

\section{DENSITY CURRENT OPERATORS} \label{append_densitycurrent}

We now discuss the current operators that enter the current correlation functions defined in Equation \ref{eqcurrcurrcorrel}. For that, let us consider the density current operator $\hat{\vec{j}}(t)$ defined as the variation of the Hamiltonian with respect to the vector potential $\vec{A}(t)$:
\begin{align*}
    \hat{\vec{j}}(t) &= -\,\frac{\delta\hat{\mathcal{H}}}{\delta\vec{A}(t)} .
\end{align*}
Introducing a vector potential in an electronic system requires a minimal substitution of the canonical momentum to preserve the gauge invariance,
$\hat{\vec{p}} \rightarrow \hat{\vec{p}} + e\vec{A}(t)$,
where $e$ is the electron charge.
This substitution introduces additional terms in the Hamiltonian that depend explicitly on $\vec{A}(t)$, from which one obtains the expression for the density current:
\begin{align}\label{eq:app_j}
    \hat{\vec{j}}(t) &= -\,\frac{e}{m}\hat{\vec{p}} - \frac{e^2}{m}\hat{n}\,\vec{A}(t),
\end{align}
where the first and second terms correspond to the paramagnetic and diamagnetic contributions, respectively.  
In this expression, $m$ denotes the effective mass and $\hat{n}$ the particle--density operator.  
Both terms contribute to the linear susceptibility, defined as the variation of the current expectation value with respect to the vector potential,
\begin{align}
    \chi_{ij}(t,t') = \frac{\delta\langle \hat{J}_i(t)\rangle}{\delta A_j(t')} .
\end{align}

On the other hand, if the variation is done with respect to the electric field, $\vec{E}(t) = -\partial_t\vec{A}(t)$ we obtain the conductivity
\begin{align}
    \sigma_{ij}(t,t') &= \dfrac{\delta\left\langle \hat{J}_i(t) \right\rangle}{\delta E_j(t')}
    .
\end{align}
The relation between the vector potential and the electric field gives way to a relation between current susceptibility and conductivity.
\begin{align} \label{eq:susc_cond_rel}
    \chi_{ij}(t,t') &= \partial_{t'}\sigma_{ij}(t,t')
    .
\end{align}

In the absence of dispersion, the response depends only on the time difference $t-t'$, allowing a Fourier representation in frequency space. Equation~(\ref{eq:susc_cond_rel}) then becomes
\begin{align} \label{eq:suscjj-cond}
    \chi_{ij}(\omega) &= \imath\omega\,\sigma_{ij}(\omega).
\end{align}

To compute the susceptibility, we can split the total current $ \hat{\vec{j}}(t)$ (Equation \ref{eq:app_j}) into paramagnetic and diamagnetic contributions. The diamagnetic term of the current is
\begin{align}
    \left\langle \hat{\vec{j}}^d(t) \right\rangle &= -\dfrac{e^2}{m} \left\langle\hat{n}\right\rangle\vec{A}(t)
    ,
\end{align}
so, this term contributes to the susceptibility with
\begin{align}
    \chi_{ij}^d(t,t') &= -\dfrac{e^2}{m}\left\langle\hat{n}\right\rangle \delta(t-t') \delta_{ij}
\end{align}
and transformed in the frequency domain as
\begin{align} \label{eq:suscjj-diam}
    \chi_{ij}^d(\omega) &= -\dfrac{e^2}{m}\left\langle\hat{n}\right\rangle \delta_{ij}
    .
\end{align}

The diamagnetic contribution $\chi^{d}_{ij}(\omega)$ is instantaneous (time-local), symmetric in the spatial indices, and time-reversal even. Consequently, it does not contribute to the antisymmetric, time-reversal-odd components of the optical response. Therefore, only paramagnetic current correlations enter the evaluation of the antisymmetric susceptibility $\chi^{as}_{ij}$ defined in Equation \ref{eq:xas}. This paramagnetic contribution arises from the perturbation $-\hat{\vec{j}}^{\,p}\!\cdot\!\vec{A}(t)$ term in the Hamiltonian. Its contribution to the susceptibility is obtained from the Kubo formula, which, written in imaginary time, can be expressed as

\begin{align} \label{eq:susc-mats}
    \chi_{ij}^p(\tau) &= -\left\langle \mathcal{T}_\tau \hat{\tilde{j}}^p_i(\tau) \hat{\tilde{j}}^p_j(0) \right\rangle
\end{align}

After analytical continuation, the imaginary part of the conductivity can be obtained from the real part through the Kramers--Kronig relations,
\begin{align}
    \Im\left[ \sigma(\omega) \right] 
    &= -\dfrac{1}{\pi} \int_{-\infty}^{\infty} d\omega' 
    \dfrac{\Re\left[ \sigma(\omega') \right]}{\omega' - \omega}.
\end{align}
Using the relation between conductivity and current susceptibility in Eq.~(\ref{eq:suscjj-cond}), the imaginary part of the conductivity can be written in terms of the real part of the paramagnetic susceptibility together with the diamagnetic contribution in Eq.~(\ref{eq:suscjj-diam}),
\begin{align}
    \Im\left[ \sigma_{ij}(\omega) \right] 
    &= \dfrac{1}{\omega} 
    \left\{
        \dfrac{e^2 \left\langle \hat{n} \right\rangle}{m} \delta_{ij} 
        - \Re\left[ \chi^p_{ij}(\omega) \right]
    \right\}.
\end{align}
In the dc limit ($\omega \to 0$) the conductivity must remain finite. 
Since the above expression contains a $1/\omega$ prefactor, the diamagnetic term must cancel the zero--frequency paramagnetic susceptibility, leading to
\begin{align}
    \Re\left[ \chi^p_{ii}(0) \right] 
    &= \dfrac{e^2 \left\langle \hat{n} \right\rangle}{m}.
\end{align}
This condition follows from gauge invariance and corresponds to the well-known $f$--sum rule for the current response.

\section{NON-LOCAL-TIME CORRELATIONS AND THE ANTISYMMETRIC SUSCEPTIBILITY} \label{append_nonlocaltimecorrels}
\subsection{Symmetry constraints and the antisymmetric response} \label{append_symconstraints}

Cubic symmetry $O_h$ forces the current response tensor to be isotropic, $\chi_{ij}(\omega)=\chi(\omega)\delta_{ij}$, which therefore suppresses any antisymmetric component of the susceptibility. Consequently, to obtain a finite antisymmetric response, we require a symmetry below $O_h$. This can be achieved, for instance, through anisotropic intersite hopping or lattice distortions.

In our calculations, we consider Jahn-Teller effects as the source of anisotropy for the hopping amplitudes. For that, we compute the self-energy generated by stabilizing Jahn-Teller distortions using the Galitzkii-Migdal approximation \cite{migdal1958interaction, galitskii1958application, onida2002electronic} (we discuss the procedure in Ref. \cite{minarro2025emergent}). We choose a Jahn-Teller distortion with $E_g$ symmetry (allowed in the vibronic coupling to $t_{2g}$ states \cite{khomskii2020orbital, bersuker2013jahn, streltsov2020jahn}), characterized by an angle $\theta$ and modify the Green functions obtained self-consistently by introducing an orbital polarization consistent with the stabilized Jahn-Teller mode

\begin{align} \label{eq:mod-green}
    \hat{\tilde{G}}_\vec{k}(\imath\omega_n) &= \left[ \hat{G}_\vec{k}^{-1}(\imath\omega_n) - \mathcal{E}_{jt} \left( \hat{\lambda}^3 \sin\theta + \hat{\lambda}^8 \cos\theta \right) \right]^{-1}
\end{align}
where $\mathcal{E}_{jt}$ is the energy of Jahn-teller mode computed using the Galitskii-Migdal formula \cite{migdal1958interaction}. To simplify the notation, we use $\hat{G}$ in the main text to refer to the orbitally-polarized Green function $\hat{\tilde{G}}$ defined by Equation \ref{eq:mod-green}. The two diagonal Gell--Mann matrices,
\begin{align}
\hat{\lambda}_3 &=
\begin{pmatrix}
1 & 0 & 0 \\
0 & -1 & 0 \\
0 & 0 & 0
\end{pmatrix},
\qquad
\hat{\lambda}_8 =
\frac{1}{\sqrt{3}}
\begin{pmatrix}
1 & 0 & 0 \\
0 & 1 & 0 \\
0 & 0 & -2
\end{pmatrix},
\end{align}
describe orthorhombic ($\hat{\lambda}_3$) and tetragonal ($\hat{\lambda}_8$) $E_g$ Jahn-Teller modes. In the $t_{2g}$ basis $(d_{yz},d_{zx},d_{xy})$, they measure the imbalance in the
orbital populations $\hat{\lambda}_3 \sim n_{yz}-n_{zx}, \hat{\lambda}_8 \sim n_{zx}+n_{yz}-2n_{xy}$.

The stabilization of a Jahn-Teller mode implies an anisotropy in the bond lengths $R^\alpha=R_0+\delta R^\alpha$ which, in turn, induce the anisotropic hopping amplitudes. 
Using Harrison scaling for Slater-Koster hopping amplitudes, $t(R)\propto R^{-3}$ \cite{harrison1989electronic,slater1954simplified}, a small distortion gives
\begin{equation} \label{app:eq:deltat}
\delta t^\alpha \equiv t^\alpha-t \simeq -3t\,\frac{\delta R^\alpha}{R_0}.
\end{equation}

We can expand conveniently the orbitally-polarized Green function into separate pure orbital and spin-orbital mixed spaces as
\begin{equation} \label{eq:orb+so}
\begin{aligned}
\hat{\tilde G}_{\vec{k}}(i\omega_n)
&=
\sum_{\alpha\in\{x,y,z\}}
\Big[
g_{\vec{k}}^\alpha(i\omega_n)
|\alpha\rangle\langle\alpha|\otimes\hat{\sigma}^0 \\
&\quad
-
\phi_{\vec{k}}^\alpha(i\omega_n)
\hat{l}^\alpha\otimes\hat{\sigma}^\alpha
\Big].
\end{aligned}
\end{equation}
where $\hat{l}^\alpha$ are orbital momentum operators and $\hat{\sigma}^\alpha$ Pauli spin matrices. Expression \ref{eq:orb+so} allows to make explicit separation between spin-independent contributions from orbital degrees of freedom (first term) and spin-orbit entanglement (second term). The orbital polarization can then be extracted from the equal-time Green function using the Matsubara identity for the density matrix $\hat{\rho} = \hat{G}(\beta)$. 
Projecting the density matrix onto the Gell--Mann generators gives
\begin{equation}
\langle \hat{\lambda}_\gamma \rangle
=
-\frac{1}{\mathcal N}\sum_{\vec{k}} g_{\vec{k}}^{\gamma}(\beta).
\end{equation}
This follows from
$\langle \hat{\lambda}_\gamma \rangle = \mathrm{Tr}[\hat{\lambda}_\gamma \hat{\rho}]$
together with $\hat{\rho} = -\hat{G}(\beta)$ and the traceless property of the Gell--Mann matrices, $\mathrm{Tr}(\hat{\lambda}_\gamma)=0$, which implies
$\langle \hat{\lambda}_\gamma \rangle = -\,\mathrm{Tr}[\hat{\lambda}_\gamma \hat{G}(\beta)]$.
The corresponding Jahn--Teller distortion amplitude is therefore
\begin{equation}
Q^\gamma
=
R_0\langle \hat{\lambda}_\gamma \rangle.
\end{equation}
where $\hat{\lambda}_\gamma = \hat{\lambda}^3 \sin\theta + \hat{\lambda}^8 \cos\theta $ denotes an orbital polarization along the unit vector
$\hat{\gamma}=(\cos\theta,\sin\theta)$ in the $(\lambda_3,\lambda_8)$ plane. By imposing volume conservation $\sum_\alpha \delta R^\alpha = 0$, the change in bond-length along the axis $\alpha$ can be written in terms of the components of orbital polarization $g_{\vec k}^\gamma(\beta)$ as

\begin{align}
\delta R^\alpha
&=
-\dfrac{2R_0}{3\mathcal{N}}
\sum_{\vec{k},\gamma}
g_\vec{k}^\gamma(\beta)
\left[
2\delta_{\alpha\gamma}-(1-\delta_{\alpha\gamma})
\right].
\end{align}

Using the Harrison scaling of the Slater–Koster hopping amplitudes, the corresponding change in the hopping amplitude becomes

\begin{align}
\delta t^\alpha
&=
\dfrac{2t}{\mathcal{N}}
\sum_{\vec{k},\gamma}
g_\vec{k}^\gamma(\beta)
\left[
2\delta_{\alpha\gamma}-(1-\delta_{\alpha\gamma})
\right].
\end{align}

Our calculations provide values for the anisotropy of the hopping amplitude in the order of $t^\alpha/t \approx 10^{-3}-10^{-2}$. Taking into account these values, we compute our lattice model assuming tetragonal elongations along $z$, where $t^x=t^y\neq t^z$ and \[\frac{t^{z}-t^{x}}{t^{x}}=\frac{t^{z}-t^{y}}{t^{y}}\approx 10^{-3}\text{--}10^{-2}.\]

\subsection{Derivation of the antisymmetric susceptibility tensor} \label{append_lderivationantisymsuscept}
Once the symmetry is reduced below $O_h$, a finite antisymmetric component is allowed in the susceptibility. As discussed in the main text, the key ingredient is the coupling to dynamical fluctuations, where the non-commutativity of transverse fluctuation operators together with time non-local corrections to the current vertex generate the finite spectral curvature. It is therefore important to understand the properties of correlations involving time non-local fermionic operators, and how these effects enable a finite curvature. Bearing this in mind, we first briefly review the formalism of non-local-time operators and, subsequently, explain why the properties of non-local matrices, particularly the cyclical properties of the trace, enable a finite, nonzero antisymmetric response.   
\\

\subsubsection{Non-local time correlations} 

Let us consider two time non-local operators $\hat{\mathcal{P}},\hat{\mathcal{Q}}$ 

\begin{subequations}
    \begin{align}
        \hat{\mathcal{P}}(\tau) &= \int d\tau' \hat{\vec{c}}^\dagger(\tau) \hat{p}(\tau-\tau') \hat{\vec{c}}(\tau') \\
        \hat{\mathcal{Q}}(\tau) &= \int d\tau' \hat{\vec{c}}^\dagger(\tau) \hat{q}(\tau-\tau') \hat{\vec{c}}(\tau')
    \end{align}
\end{subequations}

whose correlation in imaginary time is expressed as

\begin{align} \label{eq:boson-corr}
    \Pi(\tau) &= \left\langle \hat{\mathcal{P}}(\tau) \hat{\mathcal{Q}}(0) \right\rangle.
\end{align}

This expression can be expanded through quantum indices
\begin{align}
\Pi(\tau) &= \sum_{\mu,\nu} \sum_{\mu',\nu'}
\iint d\tau' d\tau'' \,
p_{\mu\nu}(\tau-\tau') q_{\mu'\nu'}(-\tau'') \notag \\
&\quad \times
\left\langle
\hat{c}^\dagger_\mu(\tau) \hat{c}_\nu(\tau')
\hat{c}_{\mu'}^\dagger(0)\hat{c}_{\nu'}(\tau'')
\right\rangle.
\label{eq:boson-corr2}
\end{align}
These correlators can be computed with two-particle Green's function
\begin{align}
\begin{split}
G^{(2)}_{\mu\nu\mu'\nu'}(\tau_1,\tau_2,\tau_3,\tau_4)
&= \left\langle \hat{\mathcal{T}}\,
\hat{c}_\mu^\dagger(\tau_1)\hat{c}_\nu(\tau_2)
\hat{c}_{\mu'}^\dagger(\tau_3)\hat{c}_{\nu'}(\tau_4)
\right\rangle \\[3pt]
&= G_{\nu\mu}(\tau_2{-}\tau_1)
   G_{\nu'\mu'}(\tau_4{-}\tau_3) \\[3pt]
&\quad -\, G_{\nu\mu'}(\tau_2{-}\tau_3)
   G_{\nu'\mu}(\tau_4{-}\tau_1) \\[3pt]
&\quad +\, \Upsilon_{\mu\nu\mu'\nu'}(\tau_1,\tau_2,\tau_3,\tau_4).
\end{split}
\end{align}
where the second term corresponds to a correction beyond the mean-field approximation represented by the first term. In this work, we consider mean-field correlations, so the last term is neglected, i.e. $\Upsilon=0$.

Time non-local matrices $\hat{p}(\tau),\hat{q}(\tau)$ in Eq(\ref{eq:boson-corr2}) are defined in $0 \leq \tau < \beta$ with $\tau > \tau'$ and $0>\tau''$. The mean-field correlation is then expressed with Green's functions as

\begin{widetext}
\begin{equation}
\begin{aligned}
\Pi(\tau)
&=
\int d\tau' \,
\mathrm{tr}\!\left[
\hat{p}(\tau-\tau') \hat{G}(\tau'-\tau)
\right]
\int d\tau'' \,
\mathrm{tr}\!\left[
\hat{q}(-\tau'') \hat{G}(\tau'')
\right]
-
\iint d\tau' d\tau'' \,
\mathrm{tr}\!\left[
\hat{p}(\tau-\tau') \hat{G}(\tau')
\hat{q}(-\tau'') \hat{G}(\tau''-\tau)
\right]
\\[5pt]
&=
\mathrm{tr}\!\left\{
\left[ \hat{p} * \hat{G} \right](\beta)
\right\}
\,
\mathrm{tr}\!\left\{
\left[ \hat{q} * \hat{G} \right](\beta)
\right\}
+
\mathrm{tr}\!\left\{
\left[ \hat{p} * \hat{G} \right](\tau)
\left[ \hat{q} * \hat{G} \right](\beta-\tau)
\right\}.
\end{aligned}
\label{eq:op-correl-time}
\end{equation}
\end{widetext}
where $\left[A * B\right](\tau)$ denotes convolution in imaginary time.

\subsubsection{Cyclical properties of the trace and finite antisymmetric susceptibility} 
The cyclical properties of the trace in non-local-time correlations are crucial to understand why a finite antisymmetric response is enabled when coupling to dynamical fluctuations is included. To illustrate this point, we take the second term in Equation \ref{eq:op-correl-time}, whose convolution is written explicitly as:

\begin{widetext}
\begin{align*}
\mathrm{tr}\!\left\{ [\hat{p} * \hat{G}](\tau) [\hat{q} * \hat{G}](\beta-\tau) \right\}
&= \mathrm{tr}\!\left[
\int_0^\beta\!\!\int_0^\beta d\tau_1 d\tau_2\,
\hat{p}(\tau-\tau_1)\hat{G}(\tau_1)
\hat{q}(\beta-\tau-\tau_2)\hat{G}(\tau_2)
\right]  \\
&= \int_0^\beta\!\!\int_0^\beta d\tau_1 d\tau_2\,
\mathrm{tr}\!\left[
\hat{G}(\tau_2)\hat{p}(\tau-\tau_1)
\hat{G}(\tau_1)\hat{q}(\beta-\tau-\tau_2)
\right]
\neq \mathrm{tr}\!\left\{
[\hat{G} * \hat{p}](\tau)
[\hat{G} * \hat{q}](\beta-\tau)
\right\}.
\end{align*}
\end{widetext}

Crucially, functions that are convolved cannot be interchanged freely. Similarly, we can apply this property to the expression of the antisymmetric susceptibility response described by Equation \ref{eq:xas}, to the following term: 
\begin{align*}
    \mathrm{tr}\left\{ \hat{A}(\tau) \left[ \hat{q} * \hat{G} \right](\beta-\tau) \right\}.
\end{align*}

In this case, there is only one convolution. Developing this expression, we find

\begin{align}\label{eq:trace}
    \begin{split}
        \mathrm{tr}\left\{ \hat{A}(\tau) \right. & \left. \left[ \hat{q} * \hat{G} \right](\beta-\tau) \right\} \\ &= \int_0^\beta d\tau'\mathrm{tr}\left[ \hat{A}(\tau) \hat{q}(\beta-\tau-\tau') \hat{G}(\tau') \right] \\ &=
        \int_0^\beta d\tau'\mathrm{tr}\left[ \hat{G}(\tau') \hat{A}(\tau) \hat{q}(\beta-\tau-\tau') \right] \\ &\neq
        \mathrm{tr}\left\{ \hat{G}(\tau) \left[ \hat{A} * \hat{q} \right](\beta-\tau) \right\}.
    \end{split}
\end{align}

This expression is a generic illustration of a single imaginary-time convolution under the trace. Equation ~\ref{eq:xas} of the main text involves terms such as 
\[
\mathrm{tr}\!\left\{
\hat{\mathscr F}_{\mathbf k}(\tau)
[\delta\hat{\mathbf j}_{\mathbf k} * \hat{\tilde G}_{\mathbf k}](-\tau)
\right\}
\]
which corresponds to the expression in Equation~\ref{eq:trace} with
$\hat A(\tau)=\hat{\mathscr F}_{\mathbf k}(\tau)$,
$\hat q=\delta\hat{\mathbf j}_{\mathbf k}$, and
$\hat G(\tau)=\hat{\tilde{G}}_{\mathbf k}(\tau)$,
and other terms like
\[
\mathrm{tr}\!\left\{
\hat{\tilde G}_{\mathbf k}(\tau)
[\delta\hat{\mathbf j}_{\mathbf k} * \hat{\mathscr F}_{\mathbf k}](-\tau)
\right\}
\]
which is obtained by exchanging $\hat{\mathscr F}_{\mathbf k}$ and $\hat{\tilde{G}}_{\mathbf k}$. Equation~\ref{eq:trace} shows that the  cyclicity of the trace reorders matrices inside the integrand, but crucially, it does not exchange imaginary times $\tau$ and $\tau'$. Therefore, the two terms related by exchanging $\hat{\mathcal{F}}_{\mathbf{k}}$ and $\hat{G}_{\mathbf{k}}$ are not algebraically identical, and therefore the antisymmetric combination in Equation~\ref{eq:xas} is not constrained to vanish, wich explains why the antisymmetric component of the susceptibility (and thus, the spectral curvature) can be nonzero.

\subsubsection{Second order expansion in the vertex correction of the antisymmetric susceptibility} 

The antisymmetric part of the paramagnetic susceptibility can be written using convolution in imaginary time as

\begin{align}
\begin{split}
{\chi}^{as}_{\alpha\beta}(\vec{k},\tau)
&= {\chi}^{p}_{\alpha\beta}(\vec{k},,\tau) - {\chi}^{p}_{\beta\alpha}(\vec{k},,\tau) \\
&= -\sum_{\vec{k}}
\mathrm{tr}\!\left\{
\left[ \hat{\Gamma}^{\alpha}_{\vec{k}} * \hat{\mathscr{G}}_{\vec{k}} \right](\tau)
\left[ \hat{\Gamma}^{\beta}_{\vec{k}} * \hat{\mathscr{G}}_{\vec{k}} \right](-\tau)
\right. \\
&\qquad\qquad \left.
-
\left[ \hat{\Gamma}^{\alpha}_{\vec{k}} * \hat{\mathscr{G}}_{\vec{k}} \right](-\tau)
\left[ \hat{\Gamma}^{\beta}_{\vec{k}} * \hat{\mathscr{G}}_{\vec{k}} \right](\tau)
\right\}.
\end{split}
\end{align}

At a given momentum $\vec{k}$, the contribution to the antisymmetric susceptibility can be written as
\begin{align}
\boldsymbol{\chi}^{as}(\vec{k},\tau)
=
-\,\mathrm{tr}
\left\{
\left[ \hat{\boldsymbol{\Gamma}}_{\vec{k}} * \hat{\mathscr{G}}_{\vec{k}} \right](\tau)
\times
\left[ \hat{\boldsymbol{\Gamma}}_{\vec{k}} * \hat{\mathscr{G}}_{\vec{k}} \right](-\tau)
\right\}.
\end{align}

We expand perturbatively in the self-energy generated by the coupling to spin fluctuations. Such a perturbation enters both through the vertex correction and the Von Neumann expansion of the Green's function $\hat{\mathscr{G}}_{\vec{k}}(i\omega_n)$. To leading order, the susceptibility vector can be expanded by decomposing the full vertex into the bare current operator and the vertex correction,
\[
\hat{\boldsymbol{\Gamma}}_{\vec{k}}
=
\hat{\vec{j}}_{\vec{k}}
+
\delta \hat{\vec{j}}_{\vec{k}} .
\]

Substituting this decomposition into the previous expression yields
\begin{widetext}
\begin{equation}
\begin{aligned}
\boldsymbol{\chi}(\vec{k},\tau)
&=
- \hat{\vec{j}}_{\vec{k}} \times \hat{\vec{j}}_{\vec{k}}\,
\mathrm{tr}\!\left\{
\hat{\mathscr{G}}_{\vec{k}}(\tau)
\hat{\mathscr{G}}_{\vec{k}}(-\tau)
\right\}
\\
&\quad
-
\mathrm{tr}\!\left\{
\left[ \delta\hat{\vec{j}}_{\vec{k}} * \hat{\mathscr{G}}_{\vec{k}} \right](\tau)
\times
\left[ \delta\hat{\vec{j}}_{\vec{k}} * \hat{\mathscr{G}}_{\vec{k}} \right](-\tau)
\right\}
\\
&\quad
-
\vec{j}_{\vec{k}} \times
\mathrm{tr}\!\left\{
\hat{\mathscr{G}}_{\vec{k}}(\tau)
\left[ \delta\hat{\vec{j}}_{\vec{k}} * \hat{\mathscr{G}}_{\vec{k}} \right](-\tau)
-
\hat{\mathscr{G}}_{\vec{k}}(-\tau)
\left[ \delta\hat{\vec{j}}_{\vec{k}} * \hat{\mathscr{G}}_{\vec{k}} \right](\tau)
\right\}.
\end{aligned}
\end{equation}
\end{widetext}

The first term vanishes identically since $\vec{j}_{\vec{k}} \times \vec{j}_{\vec{k}} = \vec{0}$, so the susceptibility vector reduces to
\begin{align}
\begin{split}
\boldsymbol{\chi}^{as}(\vec{k},\tau)
&=
-
\mathrm{tr}\!\left\{
\left[ \delta\hat{\vec{j}}_{\vec{k}} * \hat{\mathscr{G}}_{\vec{k}} \right](\tau) \right.
\times
\left.\left[ \delta\hat{\vec{j}}_{\vec{k}} * \hat{\mathscr{G}}_{\vec{k}} \right](-\tau)
\right\}
\\
&\quad
- \vec{j}_{\vec{k}} \times
\mathrm{tr}\!\left\{
\hat{\mathscr{G}}_{\vec{k}}(\tau)
\left[ \delta\hat{\vec{j}}_{\vec{k}} * \hat{\mathscr{G}}_{\vec{k}} \right](-\tau)
\right.
\\
&\quad\left.
-
\hat{\mathscr{G}}_{\vec{k}}(-\tau)
\left[ \delta\hat{\vec{j}}_{\vec{k}} * \hat{\mathscr{G}}_{\vec{k}} \right](\tau)
\right\}.
\end{split}
\end{align}

Furthermore, odd orders in the perturbative expansion are traceless due to the matrix structure: for every term inside the trace, the corresponding operators are Hermitian and odd under time-reversal symmetry, and matrices with both properties must have zero trace. Consequently, the leading contribution arises at second order. After defining
\[
\hat{\mathscr{F}}_{\vec{k}}(i\omega_n)
=
\hat{\tilde{G}}_{\vec{k}}(i\omega_n)
\hat{\Sigma}^s_{\vec{k}}(i\omega_n)
\hat{\tilde{G}}_{\vec{k}}(i\omega_n),
\]

we arrive at

\begin{widetext}
\begin{equation}\label{eq:xas2}
\begin{split}
\boldsymbol{\chi}^{as, (2)}(\vec{k},\tau)
&= -\mathrm{tr} \left\{ \left[ \delta\hat{\vec{j}}_\vec{k} * \hat{G}_\vec{k} \right](\tau) \times \left[ \delta\hat{\vec{j}}_\vec{k} * \hat{G}_\vec{k} \right](-\tau) \right\} \\ &\phantom{=} -\vec{j}_\vec{k} \times \mathrm{tr}\!\Big\{
\hat{\mathscr{F}}_\vec{k}(\tau)\!\left[ \delta\hat{\vec{j}}_\vec{k} * \hat{\tilde{G}}_\vec{k} \right](-\tau)
- \hat{\mathscr{F}}_\vec{k}(-\tau)\!\left[ \delta\hat{\vec{j}}_\vec{k} * \hat{\tilde{G}}_\vec{k} \right](\tau) \\ &\phantom{=}\hspace{42pt}
+ \hat{\tilde{G}}_\vec{k}(\tau)\!\left[ \delta\hat{\vec{j}}_\vec{k} * \hat{\mathscr{F}}_\vec{k} \right](-\tau)
- \hat{\tilde{G}}_\vec{k}(-\tau)\!\left[ \delta\hat{\vec{j}}_\vec{k} * \hat{\mathscr{F}}_\vec{k} \right](\tau)
\Big\}.
\end{split}
\end{equation}
\end{widetext}

which corresponds to Equation \ref{eq:xas} of the main text. This Equation therefore implies that (in addition to non-commutative transverse quantum fluctuations) non-local time correlations are an essential ingredient to allow non-zero antisymmetric susceptibility and explain the emergent dynamical curvature.

\onecolumngrid

\section{ISOLATION OF ANTISYMMETRIC RIXS CHANNEL WITH VANISHING BARE GEOMETRIC CONTRIBUTIONS}\label{append_polarizationasymchannel}

In general, even ignoring the coupling to fluctuations, the antisymmetric RIXS scattering channel can be nonzero in a $\mathcal{P}-\mathcal{T}$ system. The reason is that the settings of the scattering experiment can contribute to the antisymmetric signal, regardless of the intrinsic properties of the material. Here we discuss how an appropriate setting of the polarization parameters can remove these unwanted contributions. As described in Section \ref{sec:signatures_RIXS}, the polarization vectors of the incident and reflected light can be written, respectively, in terms of Bloch sphere angles $\theta_i$, $\theta_f$ and $\phi_i$, $\phi_f$, while the angles of incidence and reflection $\alpha_i$ and $\alpha_f$ can be written in terms of parameters $\delta$ and $\gamma$, which allows to write the transition operators as
\begin{align}
\hat{\Lambda}
&= \frac{N_n^2}{5}
\Bigg\{
\left[
\cos\frac{\theta_i}{2}\cos\frac{\theta_f}{2}
-
\sin\frac{\theta_i}{2}\sin\frac{\theta_f}{2}\cos(2\gamma)
\right]\mathbb{I}_3
\nonumber\\
&\qquad
+
\left(
\begin{matrix}
-\cos\frac{\theta_i}{2}\cos\frac{\theta_f}{2}
&
-\mathrm{e}^{\imath\phi_i}\sin\frac{\theta_i}{2}\cos\frac{\theta_f}{2}\cos(\gamma+\delta)
\\[4pt]
\mathrm{e}^{-\imath\phi_f}\cos\frac{\theta_i}{2}\sin\frac{\theta_f}{2}\cos(\gamma-\delta)
&
\frac{1}{2}\mathrm{e}^{\imath(\phi_i-\phi_f)}\sin\frac{\theta_i}{2}\sin\frac{\theta_f}{2}
[\cos(2\gamma)+\cos(2\delta)]
\\[4pt]
-\mathrm{e}^{-\imath\phi_f}\cos\frac{\theta_i}{2}\sin\frac{\theta_f}{2}\sin(\gamma-\delta)
&
\frac{1}{2}\mathrm{e}^{\imath(\phi_i-\phi_f)}\sin\frac{\theta_i}{2}\sin\frac{\theta_f}{2}
[\sin(2\gamma)-\sin(2\delta)]
\end{matrix}
\right.
\nonumber\\
&\qquad\qquad\qquad
\left.
\begin{matrix}
-\mathrm{e}^{\imath\phi_i}\sin\frac{\theta_i}{2}\cos\frac{\theta_f}{2}\sin(\gamma+\delta)
\\[4pt]
-\frac{1}{2}\mathrm{e}^{\imath(\phi_i-\phi_f)}\sin\frac{\theta_i}{2}\sin\frac{\theta_f}{2}
[\sin(2\gamma)+\sin(2\delta)]
\\[4pt]
\frac{1}{2}\mathrm{e}^{\imath(\phi_i-\phi_f)}\sin\frac{\theta_i}{2}\sin\frac{\theta_f}{2}
[\cos(2\gamma)-\cos(2\delta)]
\end{matrix}
\right)
\Bigg\}.
\end{align}
where $\mathbb{I}_3$ is the $3\times3$ identity matrix. We model the RIXS scattering response according to the function defined in Equation \ref{eq:ica-corr-qmain}. In the following, we ignore the effect of spin fluctuations, since we want to find the polarization parameters that suppress extrinsic contributions to the RIXS scattering function. We therefore have to consider bare Green's functions $\hat{G}_{\vec{k}}(\tau)$ instead of dressed propagators $\hat{\mathscr{G}}_{\vec{k}}(\tau)$. To calculate the scattering function, we need to consider matrix operator products such as
\begin{align} \label{eq:lambdaG_at_k}
\hat{\Lambda}\hat{G}_{\vec{k}}(\tau)
=
\begin{pmatrix}
\Lambda_{xx}\hat{g}^x + \imath\Lambda_{xy}\hat{\phi}^z - \imath\Lambda_{xz}\hat{\phi}^y
&
-\imath\Lambda_{xx}\hat{\phi}^z + \Lambda_{xy}\hat{g}^y + \imath\Lambda_{xz}\hat{\phi}^x
&
\imath\Lambda_{xx}\hat{\phi}^y - \imath\Lambda_{xy}\hat{\phi}^x + \Lambda_{xz}\hat{g}^z
\\[4pt]
\Lambda_{yx}\hat{g}^x + \imath\Lambda_{yy}\hat{\phi}^z - \imath\Lambda_{yz}\hat{\phi}^y
&
-\imath\Lambda_{yx}\hat{\phi}^z + \Lambda_{yy}\hat{g}^y + \imath\Lambda_{yz}\hat{\phi}^x
&
\imath\Lambda_{yx}\hat{\phi}^y - \imath\Lambda_{yy}\hat{\phi}^x + \Lambda_{yz}\hat{g}^z
\\[4pt]
\Lambda_{zx}\hat{g}^x + \imath\Lambda_{zy}\hat{\phi}^z - \imath\Lambda_{zz}\hat{\phi}^y
&
-\imath\Lambda_{zx}\hat{\phi}^z + \Lambda_{zy}\hat{g}^y + \imath\Lambda_{zz}\hat{\phi}^x
&
\imath\Lambda_{zx}\hat{\phi}^y - \imath\Lambda_{zy}\hat{\phi}^x + \Lambda_{zz}\hat{g}^z
\end{pmatrix}.
\end{align}
where the Green matrix functions are divided into diagonal $\hat{g}^i$ and off-diagonal $\hat{\phi}^i$ matrices \cite{minarro2025emergent}, so according to Eq(\ref{eq:orb+so}) $\hat{g}^i = g^i \hat{\sigma}^0$ and $\hat{\phi}^i = \phi^i\hat{\sigma}^i$ , where $i=x,y,z$ is also used as a short notation for $x\equiv\lvert yz\rangle$, $y\equiv\lvert xz\rangle$ and $z\equiv\lvert xy\rangle$ and $\hat{\sigma}^0$ is the 2-dimensional identity matrix. Similarly, for terms involving momentum transfer $\vec{q}$, we have

\begin{align}  \label{eq:lambdaG_at_k+q}
\hat{\Lambda}^\dagger \hat{G}_{\vec{k}+\vec{q}}(-\tau)
=
\begin{pmatrix}
\Lambda_{xx}^* \hat{\bar{g}}^x + \imath\Lambda_{yx}^*\hat{\bar{\phi}}^z - \imath\Lambda_{zx}^*\hat{\bar{\phi}}^y
&
-\imath\Lambda_{xx}^*\hat{\bar{\phi}}^z + \Lambda_{yx}^*\hat{\bar{g}}^y + \imath\Lambda_{zx}^*\hat{\bar{\phi}}^x
&
\imath\Lambda_{xx}^*\hat{\bar{\phi}}^y - \imath\Lambda_{yx}^*\hat{\bar{\phi}}^x + \Lambda_{zx}^*\hat{\bar{g}}^z
\\[4pt]
\Lambda_{xy}^* \hat{\bar{g}}^x + \imath\Lambda_{yy}^*\hat{\bar{\phi}}^z - \imath\Lambda_{zy}^*\hat{\bar{\phi}}^y
&
-\imath\Lambda_{xy}^*\hat{\bar{\phi}}^z + \Lambda_{yy}^*\hat{\bar{g}}^y + \imath\Lambda_{zy}^*\hat{\bar{\phi}}^x
&
\imath\Lambda_{xy}^*\hat{\bar{\phi}}^y - \imath\Lambda_{yy}^*\hat{\bar{\phi}}^x + \Lambda_{zy}^*\hat{\bar{g}}^z
\\[4pt]
\Lambda_{xz}^* \hat{\bar{g}}^x + \imath\Lambda_{yz}^*\hat{\bar{\phi}}^z - \imath\Lambda_{zz}^*\hat{\bar{\phi}}^y
&
-\imath\Lambda_{xz}^*\hat{\bar{\phi}}^z + \Lambda_{yz}^*\hat{\bar{g}}^y + \imath\Lambda_{zz}^*\hat{\bar{\phi}}^x
&
\imath\Lambda_{xz}^*\hat{\bar{\phi}}^y - \imath\Lambda_{yz}^*\hat{\bar{\phi}}^x + \Lambda_{zz}^*\hat{\bar{g}}^z
\end{pmatrix}.
\end{align}

where barred Green function components $\hat{\bar{g}}^i$ $\hat{\bar{\phi}}^i$ in \ref{eq:lambdaG_at_k+q} are evaluated at $(\vec{k+q}, -\tau)$, while unbarred components $\hat{{g}}^i$ $\hat{{\phi}}^i$ in Equation \ref{eq:lambdaG_at_k} are evaluated at $(\vec{k}, \tau)$. This allows us to write the following trace, necessary to calculate the scattering function in Equation \ref{eq:scattering_funct_with_fluctuations}:

\begin{align}
\frac{1}{2}\mathrm{tr}\!\left[
\hat{\Lambda}\hat{\tilde{G}}_{\vec{k}}(\tau)
\hat{\Lambda}^\dagger\hat{\tilde{G}}_{\vec{k}+\vec{q}}(-\tau)
\right]
&=
|\Lambda_{xx}|^2 g^x \bar{g}^x
-\Lambda_{xy}\Lambda_{yx}^* \phi^z \bar{\phi}^z
-\Lambda_{xz}\Lambda_{zx}^* \phi^y \bar{\phi}^y
+\Lambda_{xx}\Lambda_{yy}^*\phi^z\bar{\phi}^z
+\Lambda_{xx}\Lambda_{zz}^*\phi^y\bar{\phi}^y
\nonumber\\
&\quad
+|\Lambda_{yy}|^2 g^y\bar{g}^y
-\Lambda_{yx}\Lambda_{xy}^* \phi^z \bar{\phi}^z
-\Lambda_{yz}\Lambda_{zy}^*\phi^x\bar{\phi}^x
+\Lambda_{yy}\Lambda_{xx}^*\phi^z\bar{\phi}^z
+\Lambda_{yy}\Lambda_{zz}^*\phi^x\bar{\phi}^x
\nonumber\\
&\quad
+|\Lambda_{zz}|^2 g^z\bar{g}^z
-\Lambda_{zx}\Lambda_{xz}^* \phi^y \bar{\phi}^y
-\Lambda_{zy}\Lambda_{yz}^*\phi^x\bar{\phi}^x
+\Lambda_{zz}\Lambda_{xx}^*\phi^y\bar{\phi}^y
+\Lambda_{zz}\Lambda_{yy}^*\phi^x\bar{\phi}^x
\nonumber\\
&\quad
+|\Lambda_{xy}|^2 g^y\bar{g}^x
+|\Lambda_{xz}|^2 g^z\bar{g}^x
+|\Lambda_{yx}|^2 g^x\bar{g}^y
+|\Lambda_{yz}|^2 g^z\bar{g}^y
+|\Lambda_{zx}|^2 g^x\bar{g}^z
+|\Lambda_{zy}|^2 g^y\bar{g}^z
\nonumber\\
&=
|\Lambda_{xx}|^2 g^x \bar{g}^x
+|\Lambda_{yy}|^2 g^y\bar{g}^y
+|\Lambda_{zz}|^2 g^z\bar{g}^z
\nonumber\\
&\quad
+2\Re\!\left[
\Lambda_{xx}\Lambda_{yy}^*\phi^z\bar{\phi}^z
+\Lambda_{zz}\Lambda_{xx}^*\phi^y\bar{\phi}^y
+\Lambda_{yy}\Lambda_{zz}^*\phi^x\bar{\phi}^x
\right]
\nonumber\\
&\quad
-2\Re\!\left[
\Lambda_{xy}\Lambda_{yx}^* \phi^z \bar{\phi}^z
+\Lambda_{zx}\Lambda_{xz}^* \phi^y \bar{\phi}^y
+\Lambda_{yz}\Lambda_{zy}^* \phi^x \bar{\phi}^x
\right]
\nonumber\\
&\quad
+|\Lambda_{xy}|^2 g^y\bar{g}^x
+|\Lambda_{xz}|^2 g^z\bar{g}^x
+|\Lambda_{yx}|^2 g^x\bar{g}^y
+|\Lambda_{yz}|^2 g^z\bar{g}^y
+|\Lambda_{zx}|^2 g^x\bar{g}^z
+|\Lambda_{zy}|^2 g^y\bar{g}^z .
\end{align}

By imposing the condition $\theta_i=\pi-\theta_f$ and $\phi=\phi_i=\phi_f-\pi$, we define the antisymmetric signal as the difference between the scattering functions where the incoming and outgoing polarization configurations are exchanged (i.e., exchanging $i \leftrightarrow f$ in the polarization sector). Keeping this in mind, it is convenient to denote the tilded operator $\hat{\tilde{\Lambda}}$ as the transition operator corresponding to the exchanged polarization channel. Note that $\mathrm{diag}\,\hat{\Lambda}=\mathrm{diag}\,\hat{\tilde{\Lambda}}^{*}$ and $\Lambda_{ab}\Lambda_{ba}^{*}=\tilde{\Lambda}_{ab}\tilde{\Lambda}_{ba}^{*}$. Using these relations, the antisymmetric scattering function (within the approximation that neglects fluctuations) can be expressed in a compact manner in terms of $\hat{\Lambda}$ and $\hat{\tilde{\Lambda}}$ as follows

\begin{align} \label{eq:scattering_funtion_no_fluctuations}
\Pi^{\mathrm{as},(0)}
&\sim
\sum_{a \neq b}
\left[
|\Lambda_{ab}|^2 - |\tilde{\Lambda}_{ab}|^2
\right] g^b \bar{g}^a
\nonumber\\
&=
\left[
\sin^2\dfrac{\theta_i}{2}\cos^2\dfrac{\theta_f}{2}
-
\cos^2\dfrac{\theta_i}{2}\sin^2\dfrac{\theta_f}{2}
\right]
\Big[
\cos(\gamma+\delta)g^y\bar{g}^x
+
\sin(\gamma+\delta)g^z\bar{g}^x
-
\cos(\gamma-\delta)g^x\bar{g}^y
-
\sin(\gamma-\delta)g^x\bar{g}^z
\Big].
\end{align}

In $\mathcal{P}-\mathcal{T}$ symmetric systems, and ignoring fluctuations, the RIXS scattering function $\Pi^{\mathrm{as},(0)}$ should vanish, except for the extrinsic effects mentioned above. In order to remove undesired contributions to the antisymmetric channel, we need to analyze the values of the parameters required to nullify $\Pi^{\mathrm{as},(0)}$. According to Equation \ref{eq:scattering_funtion_no_fluctuations}, this condition is satisfied when

\begin{align*}
    \sin^2\dfrac{\theta_i}{2}\cos^2\dfrac{\theta_f}{2} - \cos^2\dfrac{\theta_i}{2}\sin^2\dfrac{\theta_f}{2} = \dfrac{\cos\theta_f - \cos\theta_i}{2} = 0
\end{align*}

Recalling that we impose $\theta_i=\pi-\theta_f$ to probe the antisymmetric scattering channel, this condition is fulfilled when $\theta_i=\theta_f=\pi/2$. Therefore, we conclude that any residual contribution from extrinsic geometric effects can be eliminated by adjusting the polarizations of the incoming and reflected beams so that $\theta_i=\pi-\theta_f=\pi/2$, regardless of the values of the Bloch sphere angle $\phi$. Consequently, using these polarization settings, the antisymmetric channel described by Equation \ref{eq:scattering_funct_with_fluctuations} directly probes the effects of fluctuations, without interference of extrinsic effects, provided that the experiments can be appropriately described by direct RIXS scattering in the ultrafast collision approximation.

\twocolumngrid

\bibliographystyle{apsrev4-2_maxauth} 
\bibliography{biblio} 

\end{document}